\documentclass[12pts]{article}
\usepackage{lineno}

\usepackage[dvips]{graphicx}
\usepackage[toc,page]{appendix}
\usepackage{graphicx}
\usepackage{amsmath}
\usepackage[varg]{txfonts}

\usepackage{longtable}

\usepackage[usenames]{color}
\usepackage{xspace}
\usepackage{siunitx}

\usepackage{natbib}
\begin{document}

\title{Spectroscopic diagnostic of halos and elves detected from space-based photometers}

\author{
  F. J. P\'erez-Invern\'on$^{1}$,
  A. Luque$^{1}$,
 F. J. Gordillo-V\'azquez$^{1}$, \\
 M. Sato$^{2}$,
 T. Ushio$^{3}$,
T. Adachi$^{4}$,
A. B. Chen$^{5}$. \\
\textit{$^{1}$Instituto de Astrof\'isica de Andaluc\'ia (IAA),} \\
   \textit{CSIC, PO Box 3004, 18080 Granada, Spain.}\\
\textit{$^{2}$Faculty of Science, Hokkaido University, } \\
   \textit{Kita-10, Nishi-8, Kita-ku, Sapporo, Hokkaido 060-0810, Japan.}\\
\textit{$^{3}$Graduate School of Engineering, Osaka University, } \\
   \textit{ 2-1 Yamadaoka, Suita, Osaka 565-0871, Japan.} \\
\textit{$^{4}$Meteorological Research Institute, Japan Meteorological Agency, } \\
   \textit{Nagamine 1-1, Tsukuba, Ibaraki 305-0052, Japan.} \\
\textit{$^{5}$Institute of Space and Plasma Sciences, National Cheng Kung University,  } \\
   \textit{Tainan, Taiwan.} \\
\footnote{Correspondence to: fjpi@iaa.es. 
Article published in Journal of Geophysical Research: Atmospheres.}
}
\date{}
\maketitle

\begin{abstract}
In this work, we develop two spectroscopic diagnostic methods to derive the peak reduced electric field in Transient Luminous Events (TLEs) from their optical signals. These methods could be used to analyze the optical signature of TLEs reported by spacecraft such as ASIM (ESA) and the future TARANIS (CNES).

As a first validation of these methods, we apply them to the predicted (synthetic) optical signatures of halos and elves, two type of TLEs, obtained from electrodynamical models. This procedure allows us to compare the inferred value of the peak reduced electric field with the value computed by halo and elve models. Afterward, we apply both methods to the analysis of optical signatures of elves and halos reported by GLIMS (JAXA) and ISUAL (NSPO) spacecraft, respectively. 

We conclude that the best emission ratios to estimate the maximum reduced electric field in halos and elves are the ratio of the Second Positive System (SPS) of N$_2$ to First Negative System (FNS) of N$_2^+$, the First Positive System (FPS) of N$_2$ to FNS of N$_2^+$ and the Lyman-Birge-Hopfield (LBH) band of N$_2$ to FNS of N$_2^+$. In the case of reduced electric fields below 150~Td, we found that the ratio of the SPS of N$_2$ to FPS of N$_2$ can also be used to reasonably estimate the value of the field. Finally, we show that the reported optical signals from elves can be treated following an inversion method in order to estimate some of the characteristics of the parent lightning.

\end{abstract}

\section{Introduction}

Tropospheric lightning produces an electromagnetic field that influences the lower ionosphere. This lightning-generated electromagnetic field can heat and accelerate ionospheric electrons, triggering a cascade of chemical reactions that results in fast optical emissions, known as Transient Luminous Events (TLEs). TLEs are an optical manifestation of the coupling between atmospheric layers. The existence of TLEs was firstly proposed by \cite{Wilson1925/PPhSocLon} in 1925. However, they were not officially discovered until 1989 by \cite{Franz1990/Sci}, who reported the detection of a sprite, a type of TLE formed by a complex structures of streamers. Since the discovery of sprites, other types of TLEs have been added to the list of these electricity phenomena in the upper atmosphere. Nowadays, the most important TLEs can be divided into sprites, halos, elves, blue jets, and giant jets \citep{Pasko2012/SSR}.

The chemical impact and optical signatures of TLEs have been widely investigated by several authors \citep{Sentman2008/JGRD/1, Gordillo-Vazquez2008/JPhD, Parra-Rojas/JGR, Parra-Rojas/JGR2015, Kuo2007/JGRA, Winkler2015/JASTP}. There have also been some ground, balloon, plane and space-based instrumentation devoted to the study of TLEs, such as the ``Imager of Sprites and Upper Atmospheric Lightning" (ISUAL) \citep{Chern2003/JASTP, hsu2017/TAOC} of the National Space Organization (NSPO), Taiwan that was in operation between May 2004 and June 2016, the ``Global LIghtning and sprite MeasurementS" (GLIMS) \citep{sato2015overview, Adachi2016/JASTP} of the Japan Aerospace Exploration Agency (JAXA) between 2012 and 2015 and the "GRAnada Sprite Spectrograph and Polarimeter" (GRASSP) \citep{Parra-Rojas2013/JGR, Passas2014/IEEE, Passas2016/APO, Gordillo-Vazquez2018/JGR} and the high-speed ground-based photometer array known as PIPER \citep{Marshall2008/ITGRS}, both of them currently in operation. Despite the valuable advance in the knowledge of TLEs in the last decades, there are still several open questions about the inception and evolution of these events or their global chemical influence in the atmosphere. For instance, we do not fully understand their relation with the parent lightning. Space-based missions such as the ``Atmosphere-Space Interactions Monitor" (ASIM) of the European Space Agency (ESA) \citep{Neubert2006/ILWS} recently launched last April 2, 2018 and the future ``Tool for the Analysis of RAdiations from lightNIngs and Sprites" (TARANIS) of the Centre National d'\'Etudes Spatiales (CNES), France \citep{Blanc2007/AdSpR} will provide new information about these events. The aim of this paper is to contribute to the scientific goals of these space missions. For this purpose, we have developed spectroscopic diagnostic methods to derive the peak reduced electric field in halos and elves. The value of the electric field determines the coupling between the troposphere, the mesosphere and the lower ionosphere produced by these TLEs.

Let us now describe the different characteristics and physical mechanisms of these two types of TLEs. Halos are produced by the quasielectrostatic field produced by lightning at altitudes between 75 and 85~km. Halos are disk-shaped optical emissions with a diameter of more than 100~km and lasting less than 10~ms \citep{Pasko1996/GeoRL, Barrington-Leigh2001/JGR, Bering2002/AdSpR,Bering2004/AdSpR,Bering2004/GeoRL,Frey2007/GeoRL, Pasko2012/SSR}. 
Elves are a consequence of the electron heating produced by the lightning-emitted electromagnetic pulses (EMP) at about 88~km of altitude and with a lateral extension of about 200~km \citep{Boeck1992/GRL,Moudry2003/JASTP, Chang2010/JGRA,Adachi2016/JASTP, van_der_Velde2016/GRL}. These types of TLEs are often observed as ring-shaped optical emissions lasting less than 1~ms.

Halos and elves emit light predominatly in the first and second positive systems of the molecular neutral nitrogen (1PS~N$_2$ and the 2PS~N$_2$, or simply FPS and SPS), the first negative system of the molecular nitrogen ion (N$_2$$^+$-1NS or simply FNS), the Meinel band of the molecular nitrogen ion (Meinel N$_2$$^+$) and the Lyman-Birge-Hopfield (LBH) band of the molecular neutral nitrogen. Some authors have used the recorded intensity ratios of these spectral bands to estimate the electric field that produces molecular excitation in air discharges. In particular, \cite{Morrill2002/GeoRL, Kuo2005/GRL, Kuo2009/JGR, Kuo2013/JGRA, Paris2005/JPhD, Adachi2006/GeoRL, Liu2006/GeoRL, Pasko2010/JGRA, Celestin2010/GeoRL, Bonaventura2011/PSST, Holder2016/PSST} based their analysis on the optical emissions from the first negative system and second positive system of molecular nitrogen at 391.4~nm and 337~nm, respectively. \cite{simek2014optical} discussed the possibility of using other spectral bands (specifically, FPS and LBH) to estimate the electric field in air discharges below 100~Td. To the best of our knowledge the spectral bands FPS and LBH have not been employed to estimate the reduced electric field in TLEs. In this work, we extend the analysis of TLEs to these spectral bands.

Spacecraft devoted to the observation of TLEs are often equipped with photometers collecting photons corresponding to certain transitions between vibrational level $v^{\prime}$ to $v^{\prime\prime}$ ($v^{\prime}$, $v^{\prime\prime}$) of the FPS(3,0) in 760~nm, SPS(0,0) in 337~nm and FNS(0,0) in 391.4~nm as well as covering the spectral (LBH) band between about 150~nm and about 280~nm. For this reason, we will focus on the spectroscopic diagnostics of halos and elves from the optical signals detected in these bands of the optical spectrum.

In this work, we develop two different diagnostic methods to extract physical information from the optical signals emitted by TLEs. The first method is based on the comparison between the measured ratio of optical signals emitted in two different spectral bands with the theoretical predictions, allowing us to estimate the reduced electric field in halos and elves. The second method is an inversion procedure useful to derive the temporal evolution of the number of photons emitted by elves from the signals recorded by space-based photometers. As a first application of these methods, we apply them to the predicted (synthetic) optical signatures of halos and elves obtained in previously developed electrodynamical models \citep{PerezInvernon2018/JGR}. Afterwards, we test the developed methods with several optical signatures of halos and elves reported respectively by ISUAL and GLIMS spacecraft. 

The organization of this paper is as follows: we present the general spectroscopic diagnostics methods in section~\ref{sect:signal}. Firstly, we apply these methods to the synthetic optical emissions of halos and elves derived with the electrodynamical models described in section~\ref{sect:electrodynamical}. Secondly, the developed diagnostic methods are applied to optical signals of halos and elves recorded by spacecraft. Results and discussion of the application of these methods to synthetic and real optical signals from halos and elves are presented in section~\ref{sec:results}. The conclusions are finally presented in section~\ref{sec:conclusions}.


\section{Optical signal treatment}
\label{sect:signal}

In this section we describe two spectroscopic diagnostic methods to analyze the optical signal emitted by halos and elves. We present in subsection~\ref{sec:opticalanalysis} a method to derive the reduced electric field in halos and elves using their optical emissions. In the case of elves, the relation between their short time duration and large spatial extension implies that photons are not neccesarily observed in the same order as they were emitted. We present in subsection~\ref{telvesignal} an inversion method to derive the temporal evolution of the emitted number of photons from the recorded optical signal. 

\subsection{Deduction of the peak reduced electric field}
\label{sec:opticalanalysis}

The aim of this section is to describe an optical diagnostic method to extract the reduced electric field from the observation of light emitted by TLEs in the lower ionosphere. We explore the possibility of using this procedure to analyze the recorded optical emissions from TLEs to be recorded by ASIM and the future TARANIS space mission.

Let us define $i(t)$ as the temporal evolution of an observed intensity at a particular wavelength or interval of wavelengths. The density of the emitting species, $N_s(t)$, can be estimated from the decay constant $A$ of the transitions that produce photons in the considered wavelength as

\begin{linenomath*}
\begin{equation}
N_s(t) = \frac{i(t)}{A}. \label{densities}
\end{equation}
\end{linenomath*}

In the case of halos, elves and sprite streamers the gas temperature is low enough to consider that the plasma is far from thermodynamic equilibrium. Hence, we can use the continuity equation of the emitting species to write their temporal production rate $S(t)$ due to electron impact  as

\begin{linenomath*}
\begin{equation}
S(t) = \frac{dN_s(t)}{dt} + A  N_s(t) + Q N_s(t) \times N - C N^{\prime}(t) + O(t), \label{production}
\end{equation}
\end{linenomath*}

where $A$ is in s$^{-1}$, $Q$ represents all the quenching rate constants by air molecules of the considered species in cm$^{-3}$s$^{-1}$ and $N$ is the density of air in cm$^{-3}$ at the emission altitude. $N^{\prime}(t)$, in $cm^{-3}$, accounts for the density of all the species that populate the species by radiative cascade with rate constants $C$, in $s^{-1}$. Finally, the term $O(t)$ in $cm^{-3}s^{-1}$ includes the rest of loss processes, such as intersystem processes or vibrational redistribution, that is usually negligible compared to quenching.

We use equations~(\ref{densities}) and~(\ref{production}) to obtain the ratios of production of two different species (1 and 2) at a fixed time $t_i$ given by $S_{12} = \frac{S_1(t_i)}{S_2(t_i)}$ in a first approach. We use the magnitude $S_{12}$ and the theoretical electric field dependent ratio of electron-impact productions of species 1 and 2, given by $\nu_{12} = \frac{\nu_1(E/N)}{\nu_2(E/N)}$, to estimate the reduced electric field (ratio or the electric field and the air density) that satisfies the equation

\begin{linenomath*}
\begin{equation}
\frac{S_1(t_i)}{S_2(t_i)} \simeq \frac{\nu_1(E/N)}{\nu_2(E/N)} . \label{equality}
\end{equation}
\end{linenomath*}

The values of $\nu_i(E/N)$ for all the considered species are calculated using BOLSIG+ for air \citep{Hagelaar2005/PSST}. The reduced electric field is usually given in Townsend units, defined as 10$^{-17}$~V~cm$^{-2}$.

This approximation assumes an electric field homogeneously distributed in space. However, halo and elve emissions are produced by an inhomogeneous electric field that varies in the scale of kilometers. We propose below a method to improve this first approach and account for the spatial distribution of the electric field.

We define the function $H\left(\frac{E^{\prime}}{N}\right) $ as the number of electrons under the influence of a reduced electric field larger than $E/N$ and weighted by the air density $N$

\begin{linenomath*}
\begin{equation}
H\left(\frac{E^{\prime}}{N}\right) = \int d\vec{r} N(\vec{r}) n_e(\vec{r}) \Theta\left( \frac{E}{N}(\vec{r}) - \frac{E^{\prime}}{N} \right), \label{nen}
\end{equation}
\end{linenomath*}

where $n_e(\vec{r})$ and $\frac{E}{N}(\vec{r})$ are, respectively, the electron density and the reduced electric field spatial distributions. The symbol $\Theta$ corresponds to the step function, being 1 if $E/N > E^{\prime}/N$ or 0 in any other case. The function defined by equation~(\ref{nen}) is monotonic and decreasing. In addition, we know that $H\left(\frac{E_{max}}{N}\right) = 0 $ by definition. Therefore, we approximate it to fisrt order as

\begin{linenomath*}
\begin{equation}
H\left(\frac{E^{\prime}}{N}\right) \simeq \alpha \left( \frac{E_{max}}{N} - \frac{E^{\prime}}{N} \right), \label{NEN_line}
\end{equation}
\end{linenomath*}

where $\alpha$ is the slope of the linear approximation.

The total excitation of species $i$ by electron impact can be written as
\begin{linenomath*}
\begin{equation}
\nu_i = \alpha \int_0^{\frac{E_{max}}{N}} d\left( \frac{E^{\prime}}{N} \right) \left| \frac{dH}{d\left( \frac{E^{\prime}}{N} \right)} \right| k_i\left( \frac{E^{\prime}}{N} \right), \label{productioni_1}
\end{equation}
\end{linenomath*}

where $k_i\left( \frac{E^{\prime}}{N} \right)$ is the electron-impact excitation of species $i$, again calculated using BOLSIG+ \citep{Hagelaar2005/PSST}.

Using the derivative of equation (\ref{NEN_line}), equation (\ref{productioni_1}) can be expressed as

\begin{linenomath*}
\begin{equation}
\nu_i = \alpha \int_0^{\frac{E_{max}}{N}} d\left( \frac{E^{\prime}}{N} \right) k_i\left( \frac{E^{\prime}}{N} \right), \label{productioni}
\end{equation}
\end{linenomath*}

and the ratio of production of two species by electron impact $S_{12} = \frac{S_1(E/N)}{S_2(E/N)}$ can be finally written as

\begin{linenomath*}
\begin{equation}
S_{12} = \frac{S_1(E/N)}{S_2(E/N)} \simeq \frac{ \int_0^{\frac{E_{max}}{N}} d\left( \frac{E^{\prime}}{N} \right) k_1\left( \frac{E^{\prime}}{N} \right)} {\int_0^{\frac{E_{max}}{N}} d\left( \frac{E^{\prime}}{N} \right) k_2\left( \frac{E^{\prime}}{N} \right)}. \label{production_ratio}
\end{equation}
\end{linenomath*}

Equation (\ref{production}) allows us to calculate species production by electron impact from observed intensities, while equation~(\ref{production_ratio}) gives the theoretical reduced electric field dependence of these productions. We can then use these two equations to estimate the maximum reduced electric field underlying halo and elve optical emissions.

\subsubsection{Accuracy of the method}
\label{sec:signalscomparison}

Let us now evaluate the usefulness of each ratio between optical emissions from different excited species to be used in order to estimate the electric field. The accuracy of the developed method is limited by the sensitivity to the electric field of the production coefficients in equation~\ref{equality}. If the ratio between the pair of considered emitted species does not depend significantly on the electric field, this method could be inaccurate. Errors introduced by measurements or by the assumed spatial distribution of the electric field (see equation~(\ref{NEN_line})) could also produce a large uncertainty in the electric field that satisfies equation~(\ref{equality}).

\begin{figure}[ht]
\centering
\includegraphics[width=12cm]{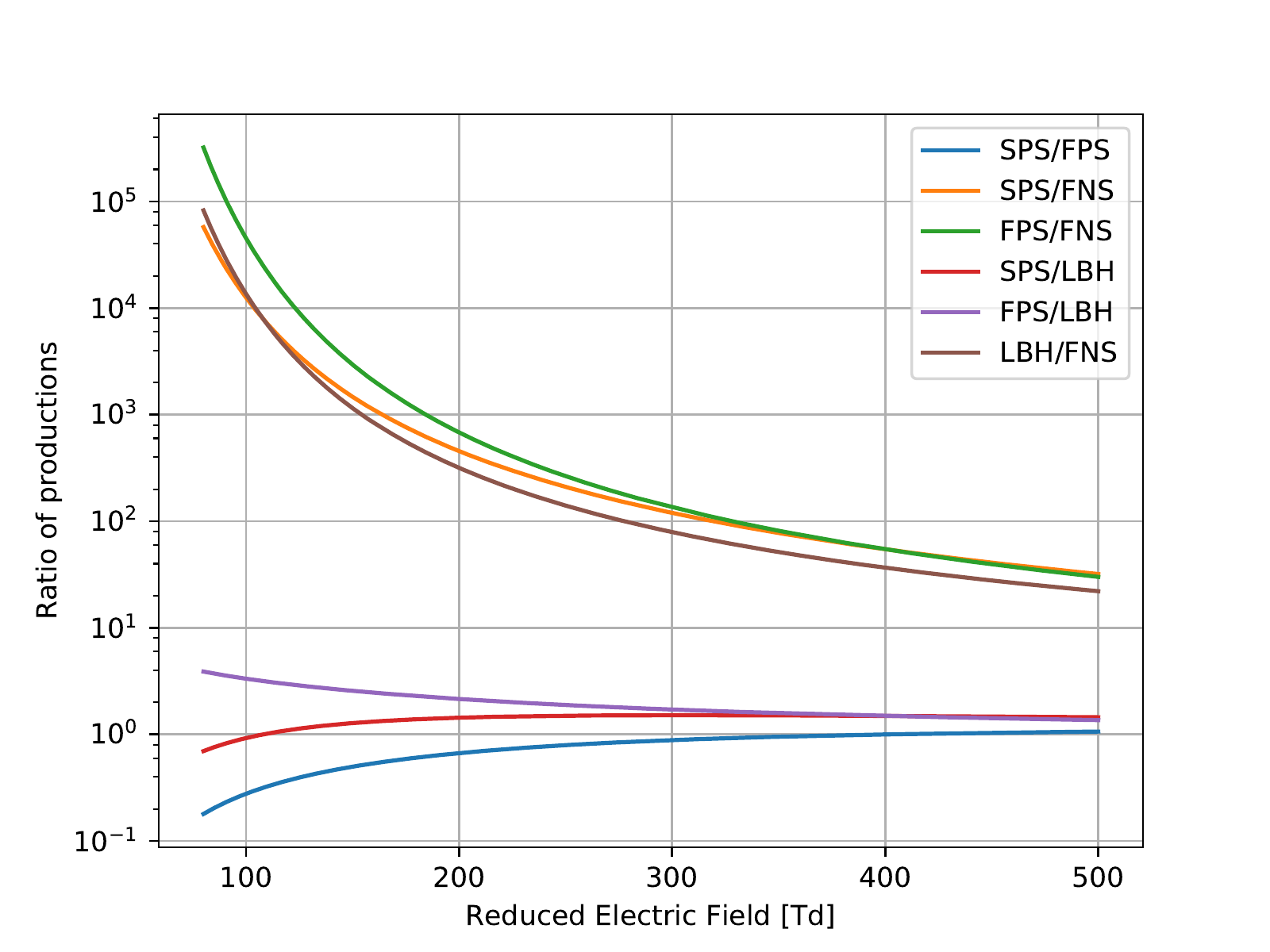}
\caption{Electric field dependence of the ratio between the rate of excitation of different states of electronically excited N$_2$. We consider excited states whose optical emissions correspond to transition within the entire First Positive System (FPS), Second Positive System (SPS), First Negative System (FNS) and Lyman-Birge-Hopfield (LBH) band. These rates have been obtained using the Boltzmann solver BOLSIG+\citep{Hagelaar2005/PSST} and the cross sections used in \cite{PerezInvernon2018/JGR}.}
\label{fig:ratio_Ered}
\end{figure}

We plot in figure~\ref{fig:ratio_Ered} the reduced electric field dependence of the ratio between several electronic emission systems of excited N$_2$. We note that the ratio of any system to the FNS depends strongly on the reduced electric field. However, the ratio of SPS to FPS, FPS to LBH and SPS to LBH do not depend significantly on the reduced electric field for high ($>$150~Td) values. Therefore, we can conclude that the ratio of FPS, SPS or LBH to FNS are the most adequate in order to estimate the reduced electric field causing optical emissions in TLEs. The ratio of FPS to SPS could also be accurate for reduced electric field values below $\sim$150-200~Td. These results are in agreement with \cite{simek2014optical}, who discussed the possibility of using different spectral bands to estimate the electric field in air discharges.

\begin{figure}[ht]
\centering
\includegraphics[width=12cm]{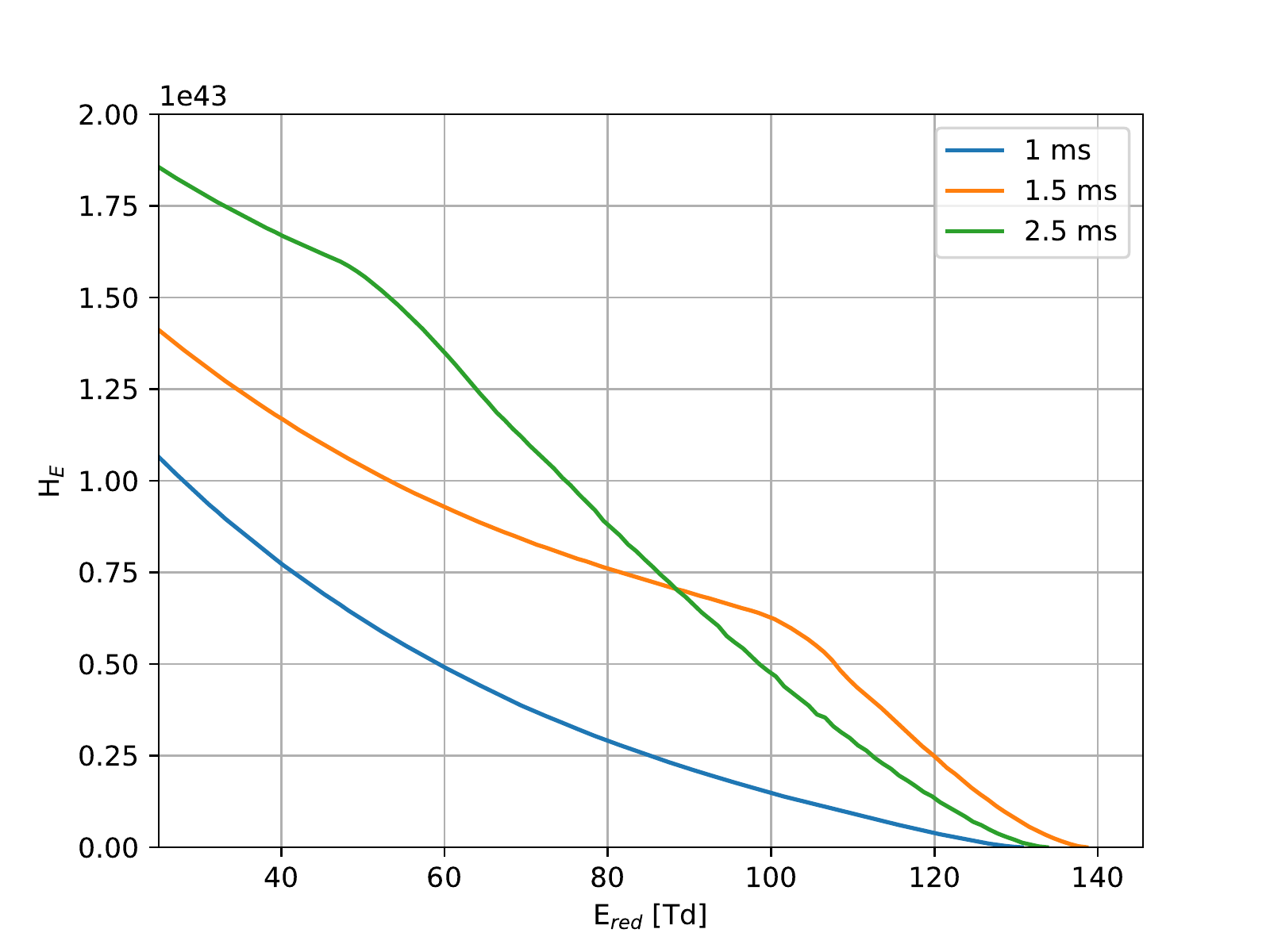}
\caption{Number of photons of electrons under the influence of a reduced electric field larger than $E/N$  and weighted by the air density. This plot corresponds to a simulation of a halo \citep{PerezInvernon2018/JGR} at different times after its onset.}
\label{fig:HE}
\end{figure}

The accuracy of the method is also determined by the goodness of the linear approximation for equation~ \ref{NEN_line}. Figure~\ref{fig:HE} clearly shows that the goodness of the linear approximation is not constant.

\subsection{Treatment of the optical signal emitted by an elve}
\label{telvesignal}

The relation between the short time (less than 1~ms) and the large spatial extension ($\sim$300~km) of elves favors an observer's simultaneous reception of photons that were not emitted at the same time from the elve. Therefore, the method developed in the previous section to derive the reduced electric field cannot be directly applied to the case of an optical signal from an elve. 

In this section, we describe an inversion method to deduce the temporal evolution of the source optical emissions of an elve knowing the signal observed by a spacecraft. Firstly, we use the source emissions of a modeled elve to calculate the observed signal as seen from a spacecraft (direct method). Then, we describe an inversion method to recover the emission source.

\subsubsection{Observed signal}
\label{observedsignal}

\begin{figure}
\centering
\includegraphics[width=12cm]{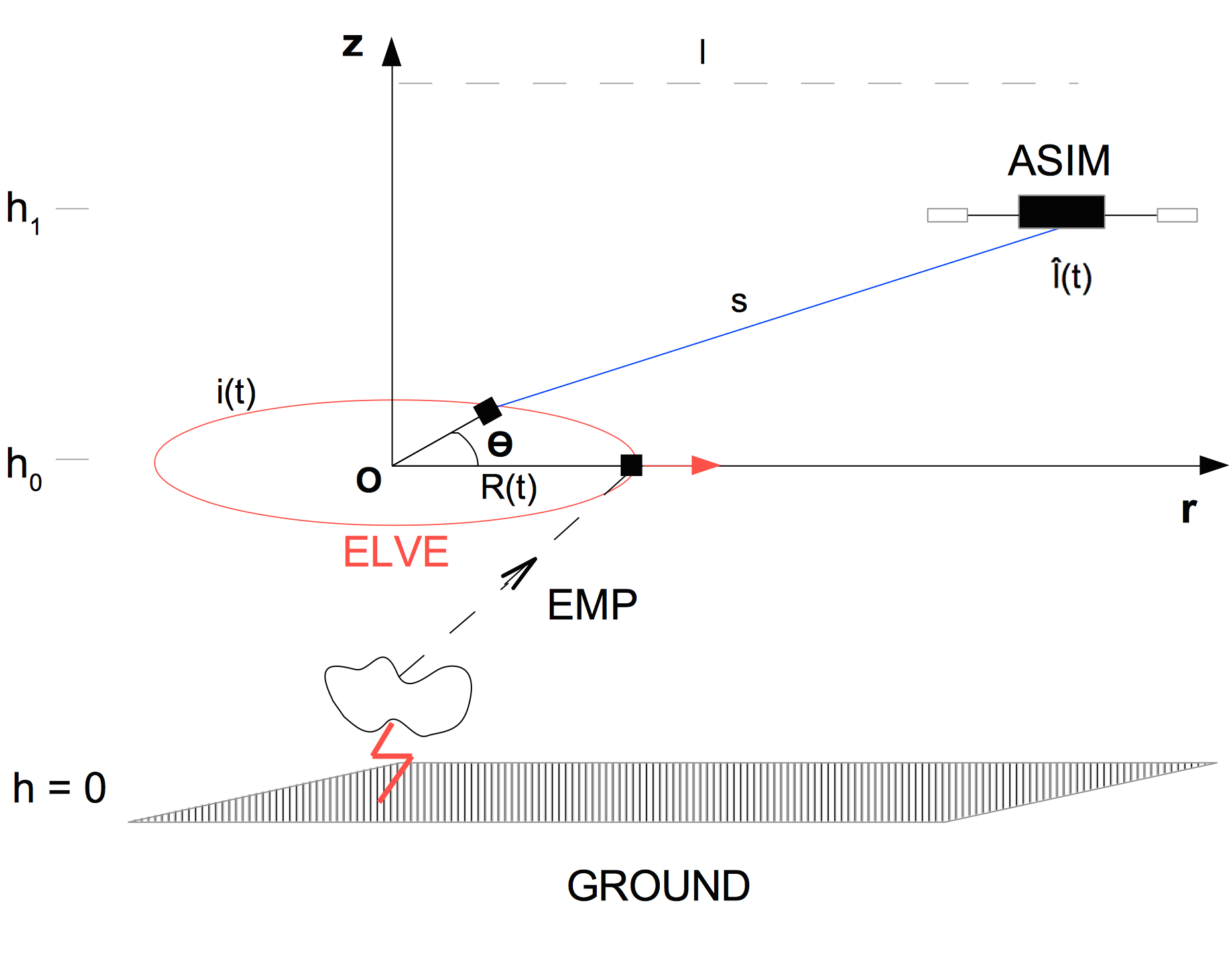}
\caption{Geometry for calculating the observed optical signal from ASIM. The elve and the spacecraft are located at altitudes $h_0$ and $h_1$ from the ground, respectively. $I(t)$ and $i(t)$ are the temporal evolution of the observed and emitted intensities, respectively. The center of the elve, denoted as $O$, is located at an horizontal distance $l$ from ASIM. $R(t)$ corresponds to the temporal dependence of the elve radius, that is radially symmetrical. The blue line ($s$) represents the path of an observed photon emitted from the elve at a point given by (R, $\theta$). }
\label{fig:geometry}
\end{figure}

The aim of this section is to describe an approximate procedure to calculate the observed signal of an elve from spacecraft given the temporal profile of emitted photons. Using a cylindrically symmetrical two-dimensional coordinate system, the elve center is located right above the lightning discharge, at an altitude $h_0$. Let's suppose that the spacecraft is located at an altitude $h_1$ and horizontally separated from the elve center by a distance $l$, as illustrated in figure~\ref{fig:geometry}. We can calculate the emitted photons per second $i(t)$ using an electrodynamical model of elves \citep{PerezInvernon2018/JGR}. It is important to note that the optical emissions are ring-shaped with a radius that increases in time according to $R(t) = (c^{2}t^{2} + 2ct(h_{0}-h_{lightning}))^{1/2}$, where $c$ is the velocity of light, the time $t$ = 0 corresponds to the zero radius of the elve and $h_{lightning}$ is the considered altitude of the parent-lightning, ranging between 0~km and the length of the lightning channel. Then, the distance $s(t)$ between an elve emitting point and the spacecraft is given by

\begin{linenomath*}
\begin{equation}
s(t) = \left[(h_1-h_0)^{2} + (l-R(t)\cos(\theta))^{2} + R^2(t)\sin^2(\theta)\right]^{1/2},  \label{s}
\end{equation}
\end{linenomath*}

where $\theta$ is the angle between the $r$-axis and the emitting point. We can now calculate the observed signal at a time $\tau$ under the approximation of isotropic emission and assuming that the elve is a thin ring. To do that, we take into account that the photons detected at a given time $\tau$ are those whose time of flight ($s(t)/c$) plus time of emissions ($t$) are equal to $\tau$

\begin{linenomath*}
\begin{equation}
\hat{I}(\tau) = \frac{A_{ph}}{4\pi} \int_{-\infty}^{\tau} i(t) R(t) dt \int_{-\pi}^{\pi} s^{-2}(t) \delta \left[ \tau - \left( t + \frac{s(t)}{c} \right) \right] d\theta , \label{Itau}
\end{equation}
\end{linenomath*}

where $A_{ph}$ is the area of the detector in the photometer.

Firstly, we calculate the angular integration, given by

\begin{linenomath*}
\begin{equation}
K(\tau, t) = \int_{-\pi}^{\pi} s^{-2}(t) \delta \left[ \tau - \left( t + \frac{s(t)}{c} \right) \right] d\theta , \label{intK}
\end{equation}
\end{linenomath*}

For this purpose, we can use the Dirac delta function property

\begin{linenomath*}
\begin{equation}
\int_{a}^{b}  f(x) \delta (G(x)) dx = \sum\limits_{i} \frac{f(x_i)}{|G^\prime(x_i)|} \label{deltaprop}
\end{equation}
\end{linenomath*}

where $x_i$ are the zeros of $G(x)$ in the interval ($a$, $b$), assuming no zeros at $a$ or $b$. In our case we have the function of $\theta$

\begin{linenomath*}
\begin{equation}
G(\theta)= \tau - \left( t + \frac{s(t)}{c} \right ) \label{G}.
\end{equation}
\end{linenomath*}

Solving $G(\theta)=0$ we obtain

\begin{linenomath*}
\begin{equation}
\theta = \pm \arccos\left( -\frac{(h_1-h_0)^{2}+l^2+R^2(t)-c^2(\tau-t)^2}{2R(t)l} \right), \label{theta}
\end{equation}
\end{linenomath*}

while the derivative of equation~(\ref{G}) is

\begin{linenomath*}
\begin{equation}
G^\prime(\theta) = -\frac{R(t)l}{c^2(\tau - t)} \sin(\theta). \label{Gpr}
\end{equation}
\end{linenomath*}

We can now combine the four last equations and replace $s \rightarrow c(\tau-t)$ to solve the angular integration under assumption of isotropic emission

\begin{linenomath*}
\begin{equation}
K(\tau,t) = \frac{2}{l(\tau-t)R(t)\sin(\theta) } . \label{kernel}
\end{equation}
\end{linenomath*}

Finally, we can replace (\ref{kernel}) in equation (\ref{Itau}) and integrate in time to obtain the observed signal. We distinguish between two possible cases:

\begin{enumerate}
\item If the center of the elve is located just below the spacecraft, the horizontal distance $l$ is equal to zero. In this particular case the integrand of equation (\ref{intK}) does not depend on the angle $\theta$, and can be analytically expressed as

\begin{linenomath*}
\begin{equation}
K(\tau, t) = 2 \pi s^{-2}(t) \delta \left[ \tau - \left( t + \frac{s(t)}{c} \right) \right]. \label{Kl0}
\end{equation}
\end{linenomath*}

As a consequence, the integration given by equation (\ref{Itau}) can be solved analytically using the Dirac's delta function properties to obtain the observed signal $\hat{I}(\tau)$.

\item In a more general case, there exists a non-zero horizontal distance $l$ between the elve center and the spacecraft, therefore the integration~(\ref{Itau}) must be numerically solved. It is important to integrate carefully over the angle $\theta$ near the singularities of $K(\tau, t)$ in equation~\ref{kernel} given by $\theta$=0, $\pi$. We refer to values of $t$ in equation~(\ref{theta}) producing these $\theta$ as $t_{inf}$ and $t_{sup}$. The value of $t_{inf}$ and $t_{sup}$ can be obtained by setting $\cos \theta = \pm 1$ in equation~(\ref{theta}) and solving for $t$, that is, 

\begin{linenomath*}
\begin{equation}
\pm 1 = \frac{(h_1-h_0)^{2}+l^2+R^2(t)-c^2(\tau-t)^2}{2R(t)l}. \label{thetasing}
\end{equation}
\end{linenomath*}

The kernel, represented by equation~(\ref{kernel}), contains integrable singularities at $t_{inf}$ and $t_{sup}$ as a consequence of the singularities of equation~(\ref{kernel}). The integration will be then performed assuming a piecewise-constant emitted intensity as

\begin{linenomath*}
\begin{equation}
\hat{I}(\tau) = \frac{A_{ph}}{4\pi} \int_{-\infty}^{\tau} K(\tau, t) i(t) dt \simeq \frac{A_{ph}}{4\pi} \sum\limits_j i_j \int_{\max(t_{inf}, t_{j-\frac{1}{2}})}^{\min(t_{sup}, t_{j+\frac{1}{2}})} K(\tau, t) dt \label{piecewise}
\end{equation}
\end{linenomath*}

\end{enumerate}

Elves are extensive structures of light with radius of more than 200~km, therefore it is possible that some emitted photons are out of the photometer field of view (FOV). Assuming a circular photometer aperture with a given FOV angle and knowing the horizontal and vertical separation between the elve and the photometer, we can calculate the maximum distance $s_0$ between an elve emitting point and the photometer as 

\begin{linenomath*}
\begin{equation}
s_0 = (h_1 - h_0) \cos^{-1} \left( \frac{FOV}{2} \right). \label{s0}
\end{equation}
\end{linenomath*}

We can then calculate the observed intensity excluding the photons that come from distances greater than $s_0$ using the Heaviside function $\Theta$. Equation (\ref{Itau}) becomes

\begin{linenomath*}
\begin{equation}
\hat{I}(\tau) = \frac{A_{ph}}{4\pi} \int_{-\infty}^{\tau} i(t) R(t) dt \int_{-\pi}^{\pi} s^{-2}(t) \delta \left[ \tau - \left( t + \frac{s(t)}{c} \right) \right] \Theta(s(t)-s_{0}) d\theta . \label{Itaus0}
\end{equation}
\end{linenomath*}

\begin{figure}
\centering
\includegraphics[width=12cm]{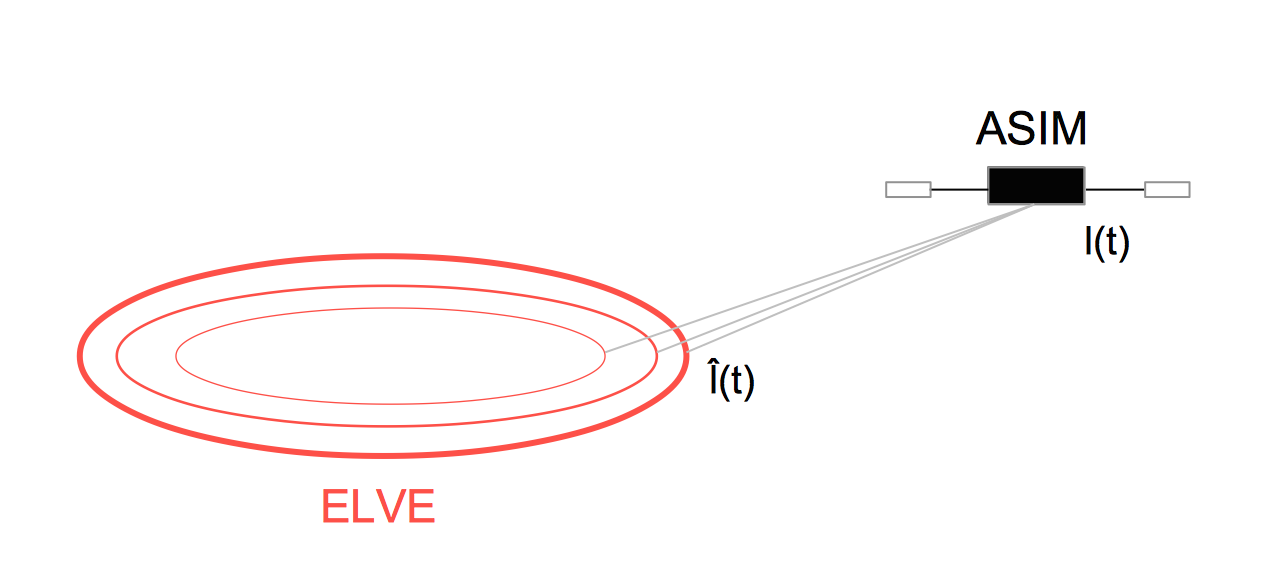}
\caption{Approximation of an elve as a succession of thin rings. $\hat{I}(t)$ corresponds to the signal observed from ASIM, while $I(t)$ would be the observed signal observed if all the emissions were focused in an instantaneous and, consequently, thin ring.}
\label{fig:rings}
\end{figure}

This method is valid if the emissions are concentrated on a thin ring. According to \cite{Rakov2003/ligh.book}, the typical rise time of CG lightning is of the order of microseconds or tens of microseconds, which corresponds to wavelengths of the order of hundreds of meters or a few of kilometers. The ionization front that causes the elve has then an approximate radius that can range between hundreds of meters and a few thousands of meters. We consider this radius negligible (compare to the elve's size) and approximate the elve ionization front as a thin ring. This approximation is justified after considering that the most impulsive CG lightning are responsible of most of the observed elves, as the absolute value of the reduced electric field in the pulse is proportional to the rise time of the discharge. However, the molecules excited by the lightning-radiated pulse do not decay instantaneously. These emitting species decay according to a radiative decay constant $\nu$. Therefore, the elve would be seen as a ring with a thickness and a radial brightness dependency determined by the radiative decay constant of each species. We can then approximate the elve as a sequence of thin rings that emit with different intensities (see figure~\ref{fig:rings}), resulting in an observed intensity $I(\tau)$ that can be calculated as the convolution of each ring intensity with its corresponding decay function as

\begin{linenomath*}
\begin{equation}
I(\tau) = \int_{0}^{\tau} \exp(-\nu t) \hat{I} (\tau - t) dt , \label{rings}
\end{equation}
\end{linenomath*}

Finally, the atmospheric absorption at each wavelength can be applied to the observed signal $I(\tau)$ in case it is necessary.

\subsubsection{Inversion of the signal}
\label{inversion}

Following our notation, the observed optical signal from a spacecraft is denoted by $I(\tau)$. In this section we describe a procedure to invert this signal and obtain the emitting source $i(t)$ that defines the elve. Firstly, we need to deconvolve the total signal$I(\tau)$to obtain an individual ring-shaped source  $\hat{I}(\tau)$  using the Wiener deconvolution in the frequency domain. We assume a signal-to-noise ratio given by

\begin{linenomath*}
\begin{equation}
SNR(t) = \frac{\sqrt{I(\tau) \Delta t}}{\Delta t} , \label{SNR}
\end{equation}
\end{linenomath*}

where $\Delta t$ is the integration time of the observed signal. Now we define the Fourier transform of the signal-to-noise ratio as

\begin{linenomath*}
\begin{equation}
SNR_f(f) = \mathcal{F}[SNR]. \label{SNR_f}
\end{equation}
\end{linenomath*}

As we explained before, the size of the ring-shaped emissions is a consequence of spatial distribution of the electric field and of the radiative decay constant ($\nu$) of the emitting species. We calculate the Fourier transform of this decay as

\begin{linenomath*}
\begin{equation}
D_f(f) = \mathcal{F}[\exp(-\nu t)], \label{decayf}
\end{equation}
\end{linenomath*}

finally, we define the Fourier transform of the observed signal as

\begin{linenomath*}
\begin{equation}
I_f(f) = \mathcal{F}[I(\tau)]. \label{If}
\end{equation}
\end{linenomath*}

We can obtain now the observed signal of each individual ring-shaped source in the frequency domain ($\hat{I}_f(f)$) using the Wiener deconvolution as

\begin{linenomath*}
\begin{equation}
\hat{I}_f(f) = \frac{I_f(f)}{D_f(f)} \left[ \frac{|D_f(f)|^2}{|D_f(f)|^2+SNR_f(f)^{-1}} \right], \label{Ihat}
\end{equation}
\end{linenomath*}

Finally, we can derive $\hat{I}(\tau)$ as the inverse Fourier transform of $\hat{I}_f(f)$ 
\begin{linenomath*}
\begin{equation}
\hat{I}(\tau) = \mathcal{F}^{-1}[I_f(f)]. \label{If}
\end{equation}
\end{linenomath*}
The next step of this inversion process is more complex and has to be accomplished numerically, since the goal is to obtain the function $i(t)$ from the integral equation (\ref{Itaus0}). The resolution of this kind of equations, known as Fredholm integral equations of the first kind, is a common problem in mathematics. We use the numerical method proposed by \cite{Hanson1971/SIAM} to solve the equation using singular values. We detail the resolution of this equation in Appendix~\ref{ap:A}.

\section{Electrodynamical models}
\label{sect:electrodynamical}

We use a halo model based on the impact of lightning-produced quasielectrostatic fields in the lower ionosphere using a cylindrically symmetrical scheme. The time evolution of the electric field is coupled with the transport of charged particles and with an extended set of chemical reactions. This model allows us to set the characteristics of the parent lightning that triggers the halo. For a complete description of the model, we refer to \cite{Luque2009/NatGe,Neubert2011/JGRA, Pasko2012/SSR, Qin2014/NatCo,Liu2015/NatCo,PerezInvernon2016/GRL, PerezInvernon2016/JGR, PerezInvernon2018/JGR}.

The model of elves is based on the resolution of the Maxwell equations and a modified Ohm's equation using a cylindrically symmetrical scheme. As in the model of halos, we can choose the characteristics of the lightning discharge that produces the elve. The details of this elve model can be found in \cite{Inan1991/GRL, Taranenko1993/GRL, Kuo2007/JGRA, Marshall2010/JGRA/2, Inan2011/Book, Luque2014/JGRA, Marshall2015/GRL, PerezInvernon2017/JGR, Liu2017/JGRAinpress, PerezInvernon2018/JGR}.

We couple the electrodynamical models of halos \citep{PerezInvernon2016/GRL} and elves \citep{PerezInvernon2016/JGR} with a set of chemical reactions collected from \cite{Gordillo-Vazquez2008/JPhD,Sentman2008/JGRD/1, Gordillo-Vazquez2009/PSST, Parra-Rojas/JGR} and \cite{Parra-Rojas/JGR2015}. We also include the molecular nitrogen vibrational kinetics proposed by \cite{Gordillo-Vazquez2010/JGRA, Luque2011/JGRA}.

The synthetic optical emissions predicted by these models in \cite{Gordillo-Vazquez2010/JGRA} and \cite{PerezInvernon2018/JGR} allow us to estimate the temporal evolution of the intensities observed by space-based photometers. Hence, we can use these predicted intensities together with the reduced electric field calculated by the electrodynamical models in order to test the accuracy of the spectroscopic diagnostic methods.

\section{Results and Discussion}
\label{sec:results}

The methods described above can be applied to the modeled optical emissions of elves and halos as well as to the optical signals recorded by spacecraft. This section is divided into two parts. In subsection~\ref{sec:signalmodels} we apply the analysis methods to the modeled optical emissions of halos and elves. This approach allows us to compare the inferred reduced electric field with the self-consistently calculated fields given by the models.

Then, we discuss in subsection~\ref{sec:missions} the possibility of applying our procedures to signals reported by ISUAL and GLIMS as well as to the future observations by ASIM and TARANIS.

\subsection{Analysis of the signals obtained with the halo and elve models}
\label{sec:signalmodels}

\subsubsection{Reduced electric field in halos}
\label{efieldhalos}

Spacecraft devoted to the observation of TLEs are often equipped with photometers collecting photons from FPS(3,0) in 760~nm, from SPS(0,0) in 337~nm and from FNS(0,0) in 391.4~nm as well as from the spectral (LBH) band between about 150~nm and about 280~nm. In this section, we discuss the possibility of using the observed optical emissions comprised in these wavelengths to deduce the reduced electric field inside a simulated halo triggered by a CG lightning discharge with a CMC of 560~C~km.

\begin{figure}[ht]
\centering
\includegraphics[width=12cm]{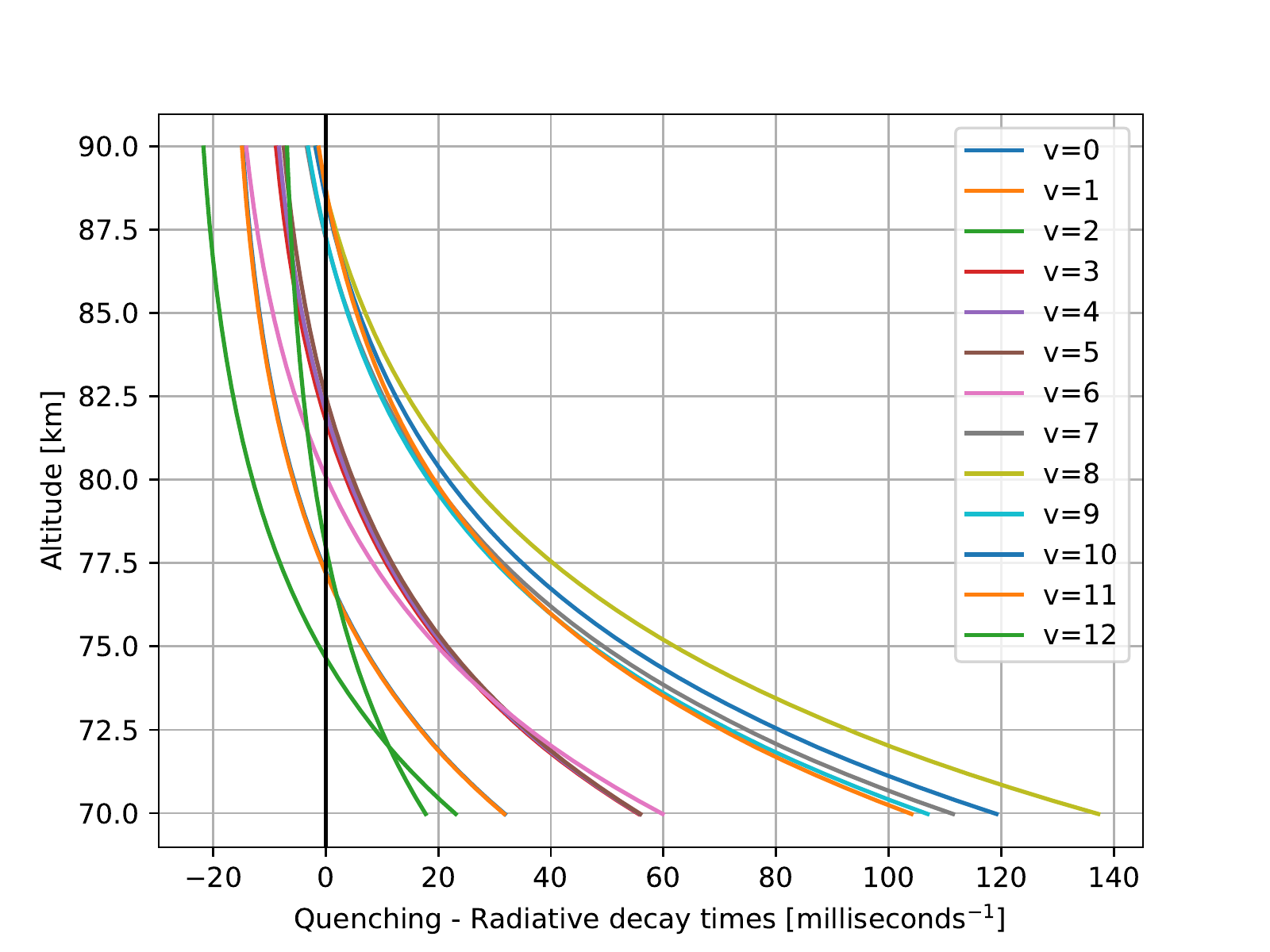}
\caption{Difference between the quenching and the radiative decay characteristic times of each vibrational level of N$_2$(a$^1$ $\Pi _g$ , v = 0, ..., 12). The altitude at which the quenching time is similar to the radiative decay time is called the quenching altitude. The quenching of each state can be neglected for altitudes significantly above the quenching altitude. The quenching of each state can be neglected for altitudes significantly above the quenching altitude. The rate constant used in this plot are from \cite{PerezInvernon2018/JGR}. }
\label{fig:quenching_altitude}
\end{figure}

The halo model allows us to obtain, among others, the temporal evolution of the optical emissions in the vibronic bands centered at 760~nm, 337~nm and 391.4~nm from the First Positive, the Second Positive Systems of N$_2$ and from the First Negative Systems of N$_2^+$. The model also computes the optical emissions in the entire LBH band, including the spectrum between 150~nm and 280~nm where space-based photometers usually record light from TLEs. We denote the intensities of these optical emissions as $I_{FPS(3,0)}(t)$, $I_{SPS(0,0)}(t)$, $I_{FNS(0,0)}(t)$ and $I_{LBH}(t)$, respectively. We can then use equations~(\ref{densities}) and~(\ref{production}) to deduce the temporal production rate of these emitting species using their kinetic rates. However, in the case of halos, the use of the observed LBH band to deduce the production rate of all the molecules emitting in these wavelengths is not possible as a consequence of their quenching rates. As halos are descending events, we cannot use a fixed altitude to estimate the quenching rates of the molecules emitting in the LBH band. In addition, we cannot neglect these quenching rates, as the quenching altitudes of some of them are located above the halo, that is, at altitudes above 80~km (see figure~\ref{fig:quenching_altitude}). Therefore, we cannot use the observed intensity of the LBH band in order to deduce the reduced electric field inside halos.

After obtaining the production rate ratios of the emitting species from their corresponding observed optical emissions, we particularize equation~(\ref{production_ratio}) to the considered emitting species to obtain the theoretical reduced electric field dependence of these ratios. To obtain these theoretical production rates, we have to include in equation~(\ref{production_ratio}) the production rates (denoted as $k$) of each species. Let us discuss the particularities of the theoretical production rate of each emitting species depending on the spectral band where the emission is produced following the kinetic scheme proposed by \cite{Gordillo-Vazquez2008/JPhD,Sentman2008/JGRD/1,Parra-Rojas/JGR, Gordillo-Vazquez2010/JGRA, Luque2011/JGRA} and \cite{Parra-Rojas/JGR2015}:

\begin{enumerate} 
\item Emissions in the vibronic band centered at 391.4~nm $\left(I_{N_2 ^+ (B^2 \Sigma ^+ _u, v = 0)}\right)$ are produced by the radiative decay process N$_2 ^+$ (B$^2$ $\Sigma ^+ _u$, v = 0) $\rightarrow$ N$_2$$^{+}$(X$^1\Sigma^+_g$, v = 0) + $h\nu$. We can estimate the number of particles of th emitting state as

\begin{linenomath*}
\begin{equation}
n_{N_2 ^+ (B^2 \Sigma ^+ _u, v = 0)} = \frac{I_{N_2 ^+ (B^2 \Sigma ^+ _u, v = 0)}}{A_{N_2 ^+ (B^2 \Sigma ^+ _u, v = 0)}}, 
\end{equation}
\end{linenomath*}

where $A_{N_2 ^+ (B^2 \Sigma ^+ _u, v = 0)}$ is the rate of the radiative decay process N$_2 ^+$ (B$^2$ $\Sigma ^+ _u$, v = 0) $\rightarrow$ N$_2$$^{+}$(X$^1\Sigma^+_g$, v = 0) + $h\nu$.
The only process that contributes to populate this state is direct electron impact ionization of N$_2$ molecules. Therefore, we can calculate the production by electron impact using equation~(\ref{production}) and exclusively considering the radiative decay process as

\begin{linenomath*}
\begin{equation}
S_{N_2 ^+ (B^2 \Sigma ^+ _u v = 0)} = \frac{dn_{N_2 ^+ (B^2 \Sigma ^+ _u v = 0)}}{dt} + A_{N_2 ^+ (B^2 \Sigma ^+ _u, v = 0)} n_{N_2 ^+ (B^2 \Sigma ^+ _u, v = 0)}
\end{equation}
\end{linenomath*}

Then, we use the rate coefficient of the reaction e + N$_2$(X$^1$ $\Sigma _g ^+$, v = 0) $\rightarrow$ e + e + N$_2 ^+$ (B$^2$ $\Sigma ^+ _u$, v = 0) in $cm^{3}s^{-1}$ to calculate the theoretical production of N$_2 ^+$ (B$^2$ $\Sigma ^+ _u$, v = 0) in equation~(\ref{productioni}).

\item Emissions in the vibronic band centered at 337~nm $\left(I_{N_2(C^3 \Pi _u , v = 0)}\right)$ are produced by the radiative decay process N$_2$(C$^3$ $\Pi _u$ , v = 0) $\rightarrow$ N$_2$(B$^3$ $\Pi _g$ , v = 0) + h$\nu$. The number of particles in the emitting state can be estimated as 

\begin{linenomath*}
\begin{equation}
n_{N_2(C^3 \Pi _u , v = 0)} = \frac{I_{N_2(C^3 \Pi _u , v = 0)}}{A_{N_2(C^3 \Pi _u, v = 0)}}, 
\end{equation}
\end{linenomath*}

where $A_{N_2(C^3 \Pi _u , v = 0)}$ is the rate of the radiative decay process N$_2$(C$^3$ $\Pi _u$ , v = 0) $\rightarrow$ N$_2$(B$^3$ $\Pi _g$ , v = 0) + h$\nu$.

There are two processes that contribute to populate the state N$_2$(C$^3$ $\Pi _u$ , v = 0), the electron impact excitation of N$_2$ and the radiative decay process N$_2$(E$^3$ $\Sigma ^+ _g$) $\rightarrow$ N$_2$(C$^3$ $\Pi _u$ , v = 0) + $h\nu$. Therefore, we include this radiative decay process in equation~(\ref{production}) to calculate the production of N$_2$(C$^3$ $\Pi _u$ , v = 0) by electron impact as

\begin{linenomath*}
\begin{equation}
\begin{split} 
S_{N_2(C^3 \Pi _u , v = 0)} = \frac{dn_{N_2(C^3 \Pi _u , v = 0)}}{dt} + A_{N_2(C^3 \Pi _u , v = 0)} n_{N_2(C^3 \Pi _u , v = 0)} - \\
A_{N_2(E^3 \Sigma ^+ _g)} n_{N_2(E^3 \Sigma ^+ _g)}
\end{split} 
\end{equation}
\end{linenomath*}

where $A_{N_2(E^3 \Sigma ^+ _g)}$ is the rate of the radiative decay process N$_2$(E$^3$ $\Sigma ^+ _g$) $\rightarrow$ N$_2$(C$^3$ $\Pi _u$ , v = 0) + $h\nu$. However, it is necessary to estimate the number of particles of N$_2$(E$^3$ $\Sigma ^+ _g$). To do that, we assume that the production rate of N$_2$(E$^3$ $\Sigma ^+ _g$) is proportional to the production rate of N$_2$(C$^3$ $\Pi _u$ , v = 0) as $S_{N_2(E^3 \Sigma ^+ _g)} = \chi S_{N_2(C^3 \Pi _u, v = 0)}$, where $\chi$ is a constant. To obtain the value of this constant, we calculate with BOLSIG+ \citep{Hagelaar2005/PSST} the rate coefficients of the reactions e + N$_2$(X$^1$ $\Sigma _g ^+$, v = 0) $\rightarrow$ e + N$_2$(C$^3$ $\Pi _u$ , v = 0) and e + N$_2$(X$^1$ $\Sigma _g ^+$, v = 0) $\rightarrow$ e + N$_2$(E$^3$ $\Sigma ^+ _g$). We obtain that the ratio between these two rate coefficients is $\chi \sim 0.02$ and does not significantly depend on the reduced electric fields.
We can then write the time derivative of $n_{N_2(E^3 \Sigma ^+ _g)}$ as

\begin{linenomath*}
\begin{equation}
\begin{split} 
\frac{dn_{N_2(E^3 \Sigma ^+ _g)}}{dt} = \chi \left( \frac{dn_{N_2(C^3 \Pi _u , v = 0)}}{dt} + A_{N_2(C^3 \Pi _u , v = 0)} n_{N_2(C^3 \Pi _u , v = 0)} \right) \\ - (\chi + 1 ) n_{N_2(E^3 \Sigma ^+ _g)}, 
\end{split} 
\end{equation}
\end{linenomath*}

that can be solved as

\begin{linenomath*}
\begin{equation}
\begin{split} 
n_{N_2(E^3 \Sigma ^+ _g)} = \chi \exp \left(-( 1 + \chi ) A_{N_2(E^3 \Sigma ^+ _g)} t \right) \int^t_0 dt^{\prime} \exp \left(-( 1 + \chi ) A_{N_2(E^3 \Sigma ^+ _g)} t^{\prime}\right) \\ \left[ \frac{dn_{N_2(C^3 \Pi _u , v = 0)}}{dt} + A_{N_2(C^3 \Pi _u , v = 0)} n_{N_2(C^3 \Pi _u , v = 0)}\right]. 
\end{split} 
\end{equation}
\end{linenomath*}

Finally, we use the rate coefficient of the reaction e + N$_2$(X$^1$ $\Sigma _g ^+$, v = 0) $\rightarrow$ e + N$_2$(C$^3$ $\Pi _u$ , v = 0) in $cm^{3}s^{-1}$ to calculate the theoretical production of N$_2$(C$^3$ $\Pi _u$ , v = 0) in equation~(\ref{productioni}). 

\item Emissions in the vibronic band centered at 760~nm $\left(I_{N_2(B^3 \Pi _g , v = 3)}\right)$ are produced by the radiative decay process N$_2$(B$^3$ $\Pi _g$ , v = 3) $\rightarrow$ N$_2$(A$^3$ $\Sigma _u ^+$ , v = 1) + $h\nu$. We can estimate the number of particles of N$_2$(B$^3$ $\Pi _g$ , v = 3) as

\begin{linenomath*}
\begin{equation}
n_{N_2(B^3 \Pi _g , v = 3)} = \frac{I_{N_2(B^3 \Pi _g , v = 3)}}{A_{N_2(B^3 \Pi _g , v = 3)}}, 
\end{equation}
\end{linenomath*}

where $A_{N_2(B^3 \Pi _g , v = 3)}$ is the rate of the radiative decay process N$_2$(B$^3$ $\Pi _g$ , v = 3) $\rightarrow$ N$_2$(A$^3$ $\Sigma _u ^+$ , v = 1) + $h\nu$.

There are several processes that contribute to populate the state N$_2$(B$^3$ $\Pi _g$ , v = 3), the electron impact excitation of N$_2$ and the radiative decay processes N$_2$(C$^3$ $\Pi _u$ , v = 0, ..., 4) $\rightarrow$ N$_2$(B$^3$ $\Pi _g$ , v = 3) + $h\nu$. Therefore, we have to include these radiative decay processes in equation~(\ref{production}) to calculate the production of N$_2$(B$^3$ $\Pi _g$ , v = 3) by electron impact as

\begin{linenomath*}
\begin{equation}
\begin{split} 
S_{N_2(B^3 \Pi _g , v = 3)} = \frac{dn_{N_2(B^3 \Pi _g , v = 3)}}{dt} + A_{N_2(B^3 \Pi _g , v = 3)} n_{N_2(B^3 \Pi _g , v = 3)} - \\
A_{N_2(C^3 \Pi _u, v = 0, ..., 4)} n_{N_2(C^3 \Pi _u , v = 0, ..., 4)},
\end{split} 
\end{equation}
\end{linenomath*}

where $A_{N_2(C^3 \Pi _u , v = 0, ..., 4)}$ are the rate of the radiative decay processes N$_2$(C$^3$ $\Pi _u$ , v = 0, ..., 4) $\rightarrow$ N$_2$(B$^3$ $\Pi _g$ , v = 3) + $h\nu$. As we have only deduced the number of species N$_2$(C$^3$ $\Pi _u$ , v = 0), we have to use the Vibrational-Distribution-Function (VDF) of the species N$_2$(C$^3$ $\Pi _u$ , v = 0, ..., 4) to estimate the number of species N$_2$(C$^3$ $\Pi _u$ , v = 1, ..., 4). The VDF of N$_2$(C$^3$ $\Pi _u$ , v = 0, ..., 4) is sensitive to the electric field \citep{simek2014optical}, specially for low fields below 200~Td. As a first approximation, we can obtain this VDF using the chemical scheme of \cite{Gordillo-Vazquez2008/JPhD,Sentman2008/JGRD/1,Parra-Rojas/JGR, Gordillo-Vazquez2010/JGRA, Luque2011/JGRA} and \cite{Parra-Rojas/JGR2015}:

\begin{linenomath*}
\begin{equation}
\label{VDFC3P}
\begin{split} 
VDF\left( N_2(C^3 \Pi _u, v = 0, ..., 4) \right) = \\ \left(0.69, 0.15, 0.12, 0.03, 0.01 \right),
\end{split} 
\end{equation}
\end{linenomath*}

where the numbers correspond to the relative population of each vibrational level. 
We can also use as an approximation to this VDF the Franck-Condon factors given in Table~25 of \cite{Gilmore1992/JPCRD}.

Then, we use the rate coefficient of the reaction e + N$_2$(X$^1$ $\Sigma _g ^+$, v = 0) $\rightarrow$ e + N$_2$(B$^3$ $\Pi _g$ , v = 3) in $cm^{3}s^{-1}$ to calculate the theoretical production of N$_2$(B$^3$ $\Pi _g$ , v = 3) in equation~(\ref{productioni}).

\end{enumerate}

\begin{figure}
\centering
\includegraphics[width=14cm]{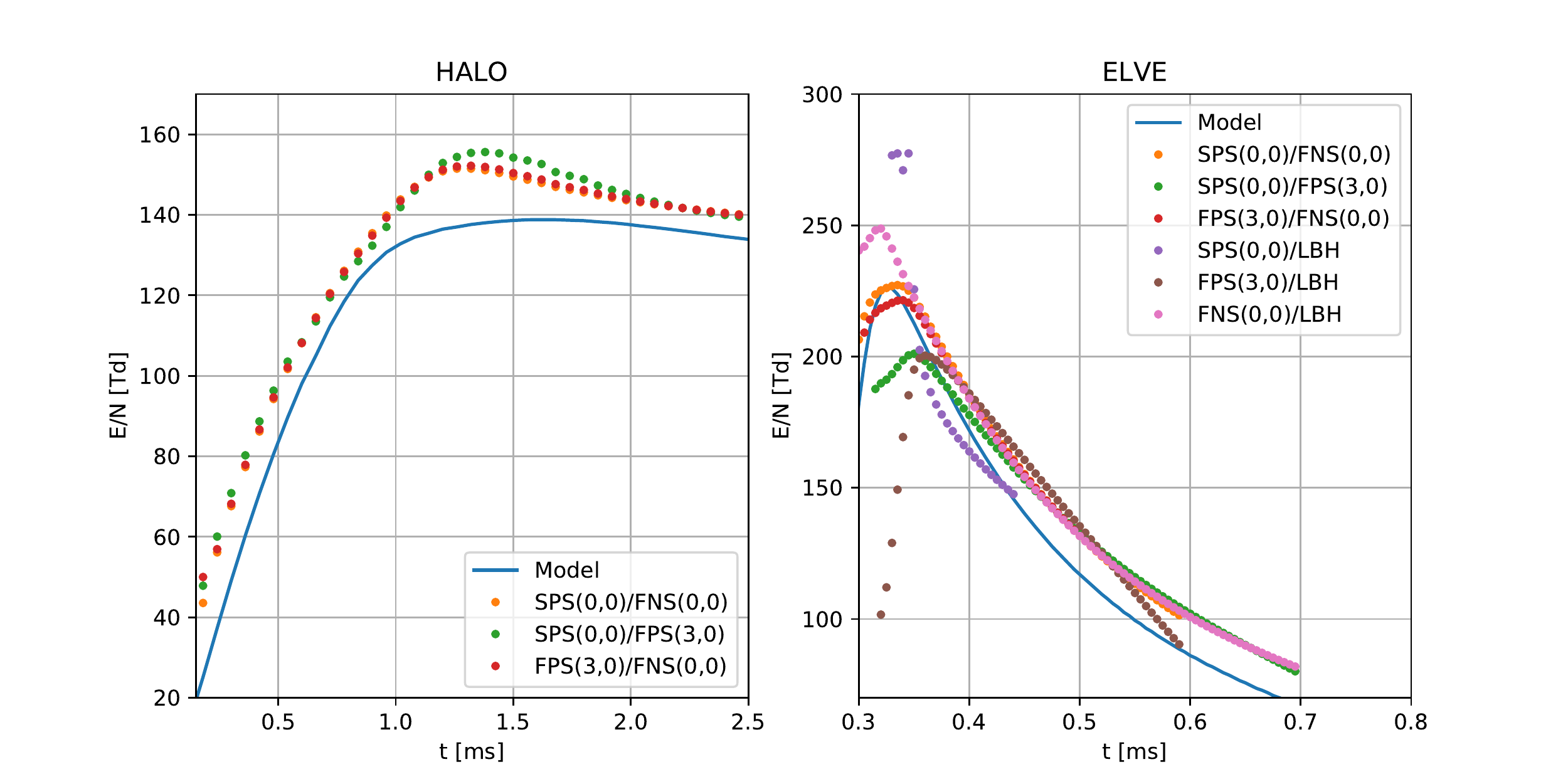}
\caption{Temporal evolution of the maximum reduced electric field inside a halo (left) and an elve (right). The blue line corresponds to the maximum reduced electric field according to the halo or elve model. The rest of the points correspond to the inferred reduced electric field using the ratios of the observed FPS(3,0), SPS(0,0), FNS(0,0) and LBH band of N$_2$ following subsection~\ref{sec:opticalanalysis}. We have used the VDF given by expression~(\ref{VDFC3P}).}
\label{fig:Estimated_Ered}
\end{figure}

Finally, we can calculate the reduced electric field necessary to match the observed and theoretical ratios of production at each time. The results are plotted in figure~\ref{fig:Estimated_Ered} together with the maximum reduced electric field given by the simulation at each particular time and using the VDF of expression~(\ref{VDFC3P}). The use of the VDF approximated as the Franck-Condon factors given in Table~25 of \cite{Gilmore1992/JPCRD} produces a different electric field (25 \% greater than the one plotted in figure~\ref{fig:Estimated_Ered}) for the case of the FPS/SPS ratio. However, it does not influence the electric field obtained from the rest of ratios. The comparison between the derived electric fields from optical band intensity ratios with the electric field given by the model allows us to test the accuracy of the proposed methods for the optical diagnosis of halos using ASIM and TARANIS optical data.

\subsubsection{Reduced electric field in elves}

Let us now apply the previous electric field deduction method to a simulated elve triggered by a lightning stroke with a current peak of 220~kA. As in the case of the halo model, the elve model allows us to calculate the intensities of the optical emissions $I_{FPS(3,0)}(t)$, $I_{SPS(0,0)}(t)$, $I_{FNS(0,0)}(t)$ and $I_{LBH}(t)$. 

Again, the first step to derive the reduced electric field inside the TLE is to use equations~(\ref{densities}) and~(\ref{production}) to deduce the temporal production rate of these emitting species using their kinetic rates. Elves are always produced at altitudes of about 88~km, where the quenching of all the vibrational states emitting in the LBH band is less important than the radiative decay (see figure~\ref{fig:quenching_altitude}). Therefore, we can now neglect in our calculations the quenching of all these species at a fixed altitude of 88~km and use the intensities observed in the LBH band to deduce the reduced electric field.

The second step is to particularize equation~(\ref{production_ratio}) to the case of the considered emitting species to obtain the theoretical reduced electric field dependence of their ratios. We use the same steps enumerated in section~\ref{efieldhalos} to deduce the theoretical production rate of each emitting species with the following exceptions:

\begin{enumerate} 
\item Emissions in the LBH band $\left(I_{N_2(a^1 \Pi _g , v = 0, ..., 15)}\right)$ are produced by the radiative decay processes N$_2$(a$^1$ $\Pi _g$ , v = 0, ..., 15) $\rightarrow$ N$_2$(X$^1$ $\Sigma _g ^+$, v = 0, ..., 8)+ $h\nu$. We can estimate the number of particles in the emitting state by neglecting the quenching at elve altitude as 

\begin{linenomath*}
\begin{equation}
n_{N_2(a^1 \Pi _g , v = 0, ..., 15)} = \frac{I_{N_2(a^1 \Pi _g , v = 0, ..., 15)}}{A_{N_2(a^1 \Pi _g , v = 0, ..., 15)}}, 
\end{equation}
\end{linenomath*}

where $A_{N_2(a^1 \Pi _g , v = 0, ..., 15)}$ is the total rate of the radiative decay processes N$_2$(a$^1$ $\Pi _g$ , v = 0, ..., 15) $\rightarrow$ N$_2$(X$^1$ $\Sigma _g ^+$, v = 0, ..., 8)+ $h\nu$.

The production of N$_2$(a$^1$ $\Pi _g$ , v = 0, ..., 15) by electron impact can be calculated using equation~(\ref{production})

\begin{linenomath*}
\begin{equation}
S_{N_2(a^1 \Pi _g , v = 0, ..., 15)} = \frac{dn_{N_2(a^1 \Pi _g , v = 0, ..., 15)}}{dt} + A_{N_2(a^1 \Pi _g , v = 0, ..., 15)} n_{N_2(a^1 \Pi _g , v = 0, ..., 15)}.
\end{equation}
\end{linenomath*}

Finally, the rate of the reaction e + N$_2$(X$^1$ $\Sigma _g ^+$, v = 0)$\rightarrow$ e + N$_2$(a$^1$ $\Pi _g$ , v = 0, ..., 15) in $cm^{-3}s^{-1}$ is used to calculate the theoretical production of N$_2$(a$^1$ $\Pi _g$ , v = 0, ..., 15) in equation~(\ref{productioni}). 

\end{enumerate}

As in the case of halos, the next step would be to calculate the reduced electric field necessary to match the observed and theoretical ratio of production at each time. The results are plotted in figure~\ref{fig:Estimated_Ered} together with the maximum reduced electric field given by the simulation. The comparison between the deduced electric fields with the electric field given by the model allows us to test the accuracy of these methods for elves. The best fits between the maximum electric field given by the model and the values deduced using the ratio of different pairs of emitted intensities is reached when the FNS is used.

\subsubsection{Emitting source of elves}

In the previous section we have deduced the reduced electric field of halos and elves considering that the observed emissions and the emitting source follow the same temporal evolution. However, as we discussed before, this assumption is not true for the case of elves. Therefore, it is necessary to invert the observed signal in order to obtain the emitting source before deducing the reduced electric field in the elve. In this section, we apply the methods described in subsection~\ref{observedsignal} to calculate how a spacecraft would observe a simulated elve optical emission. Afterwards, we invert this signal following the process detailed in subsection~\ref{inversion} to recover the emitting source. 

\begin{figure}
\centering
\includegraphics[width=13cm]{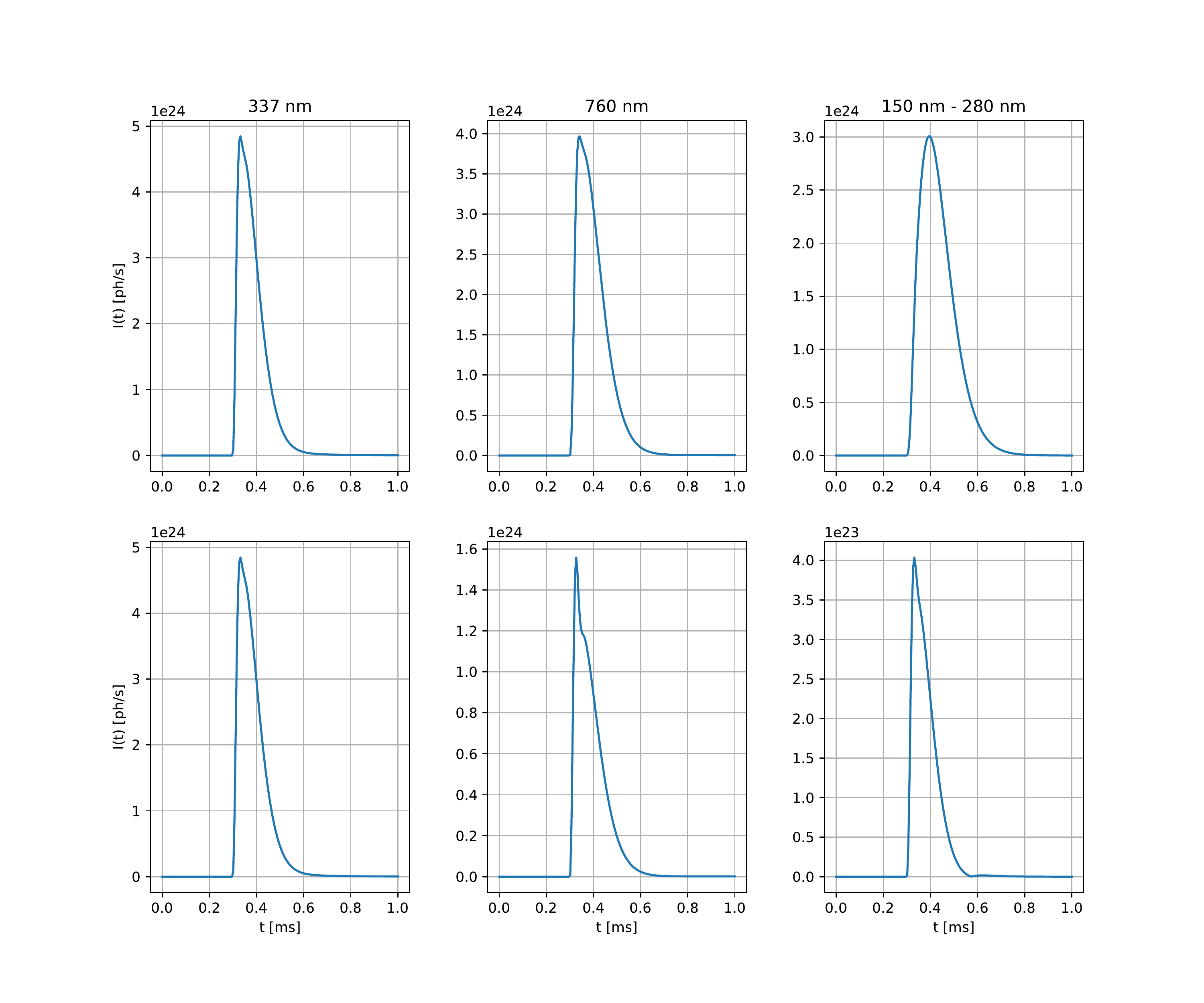}
\caption{Optical emissions in different wavelengths of a simulated elve and triggered by a lightning with a current peak of 154~kA (first row). Optical signal convolved with their corresponding decay function (second row).}
\label{fig:Source_originalandoneringTARANIS}
\end{figure}

We use as source of the optical emissions a simulated elve triggered by a lightning stroke with a current peak of 154~kA. We plot in the first row of figure~\ref{fig:Source_originalandoneringTARANIS} the emitting source that we will treat in this section.

The method developed in subsection~\ref{observedsignal} to calculate the signal observed by a spacecraft receives as input the optical emissions of a thin ring-shaped elve. We convolve the emissions shown in the first row of figure~\ref{fig:Source_originalandoneringTARANIS} with their corresponding decay function (see subsection~\ref{observedsignal}) to obtain the observed emissions due to an instantaneous and thin ring (second row of figure~\ref{fig:Source_originalandoneringTARANIS}).

\begin{figure}
\centering
\includegraphics[width=14cm]{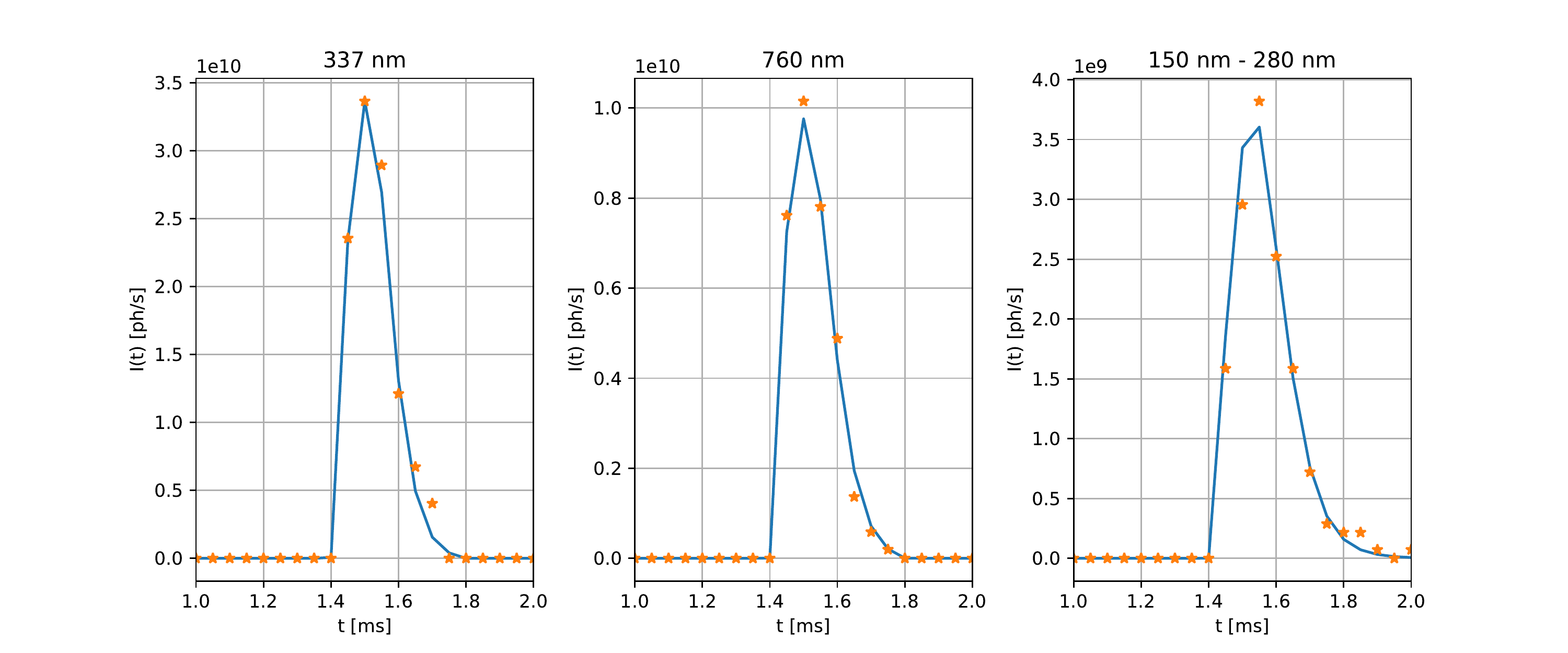}
\includegraphics[width=14cm]{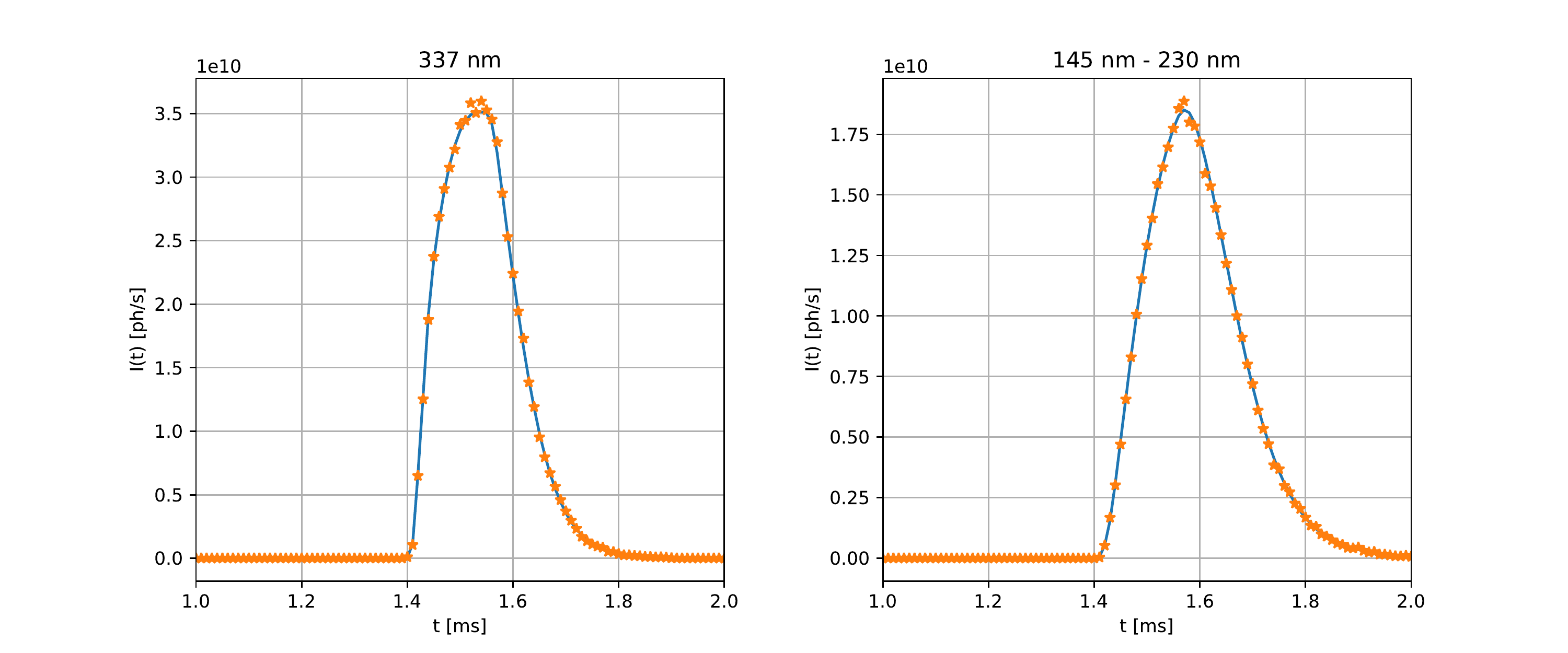}
\caption{Hypothetical signals from an elve received by photometers on-board (upper row) TARANIS and (lower row) ASIM. We assume that both spacecraft are located at an altitude of 410~km and at a horizontal distance from the center of the elve of 80~km. The blue lines correspond to the signals without noise, while the orange asterisk represents the signals with noise. We have assumed that the photometers have a FOV of 55$^{\circ}$ for the case of TARANIS and 61.4$^{\circ}$ for ASIM. The sampling frequencies are 20~kHz for TARANIS and 100~kHz for ASIM, while the circular aperture total area is 0.04~m$^{-2}$}
\label{fig:Received_TARANIS_noise}
\end{figure}

Let us now follow the method of subsection~\ref{observedsignal} to calculate the hypothetical signals observed from TARANIS and ASIM. We have assumed that the observation instruments are located at an altitude of 410~km and at a horizontal distance from the center of the elve of 80~km. We plot in figure~\ref{fig:Received_TARANIS_noise} the calculated received signals.

The inversion method described in subsection~\ref{inversion} can be directly applied to the received signals in order to recover the emitting sources. However, before applying the inversion method, we turn to a more realistic case adding some artificial noise to the received signals as follows.
We denote as $S_i$ the $i$ nth point of a received signal and as $S_{max}$ the maximum value of the signal. Then, we define a parameter $g$ that will control the noise:

\begin{linenomath*}
\begin{equation}
g =\frac{a}{S_{max}}, \label{g}
\end{equation}
\end{linenomath*}

where $a$ is an arbitrary number that controls the noise. Afterwards, we use this parameter ($g$) to generate a random number $b_i$ with a Poisson distribution with mean value $gS_i$. Finally, the $S_i$ value of the signal is modified to be

\begin{linenomath*}
\begin{equation}
S_i^{\prime} =\frac{b_i}{g}. \label{Sip}
\end{equation}
\end{linenomath*}

Setting the $a$ parameter to 50 for the case of TARANIS and 5$\times$10$^3$ for ASIM, we obtain the signal with the noise shown in figure~\ref{fig:Received_TARANIS_noise} (orange asterisk).

\begin{figure}
\centering
\includegraphics[width=14cm]{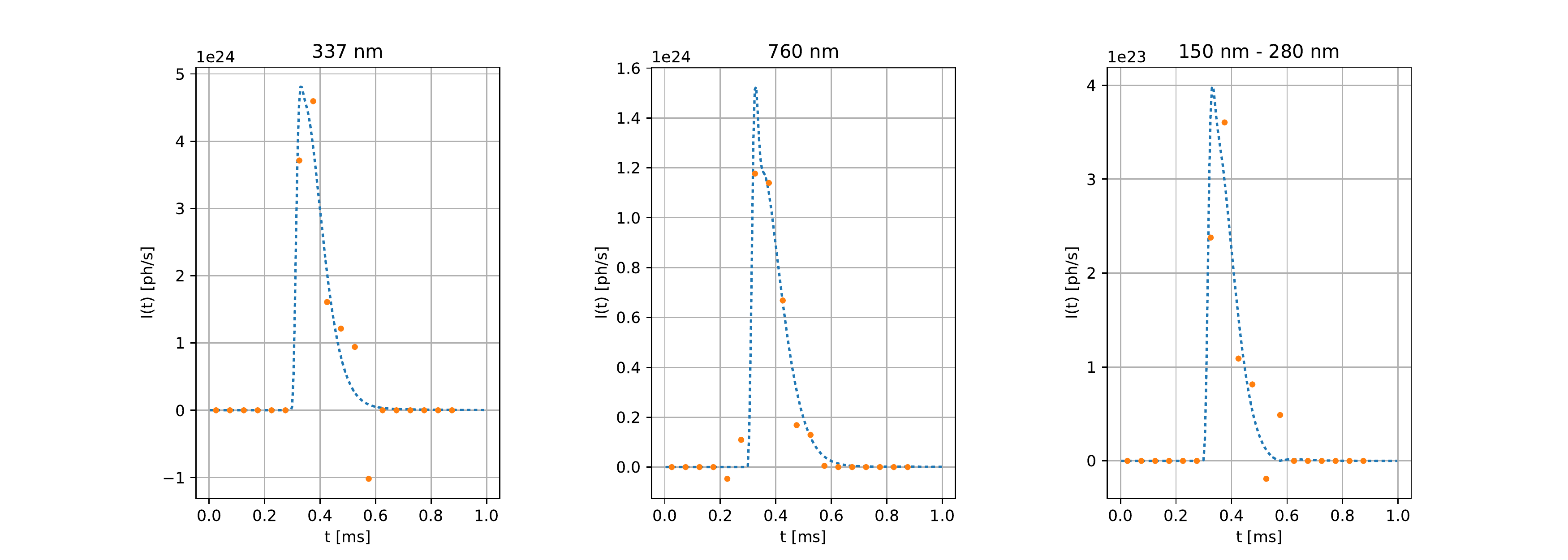}
\includegraphics[width=14cm]{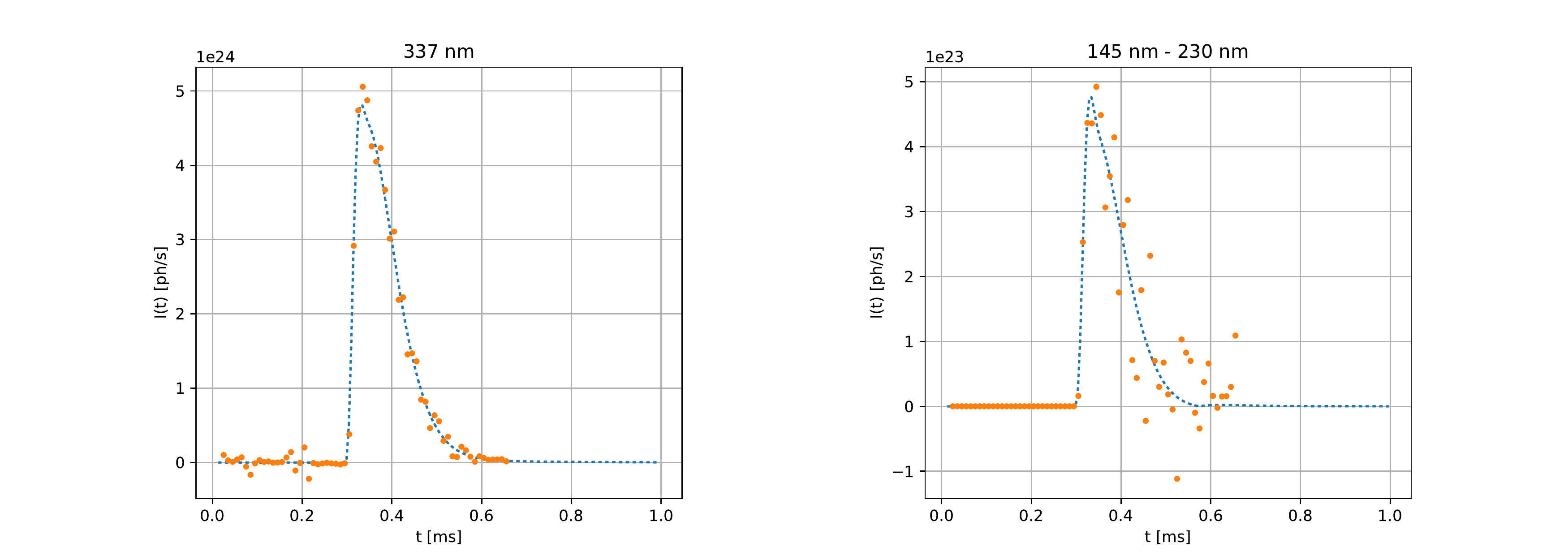}
\caption{Result of inverting the signals plotted in figure~\ref{fig:Received_TARANIS_noise} (orange asterisk). We also plot the emitted signal given by the FDTD elve model (blue line). }
\label{fig:Source_Hanson_TARANIS}
\end{figure}

We can now apply the inversion method developed in subsection~\ref{inversion} to the received signals with noise in order to compare with the signals given by the models. We plot the results in figure~\ref{fig:Source_Hanson_TARANIS}.

\subsection{Analysis of signals recorded from space}
\label{sec:missions}

In this section we analyze the particularities of the photometers integrated in ISUAL, GLIMS, ASIM and TARANIS (see table~\ref{tab:missions}) for the investigation of TLEs. In addition, we apply the inversion method described in subsection~\ref{inversion} to one LBH signal emitted by an elve detected by GLIMS.

\tiny

\begin{longtable}{|c|c|c|c|c|c|}

\hline \multicolumn{1}{|c}{\textbf{Mission}} & \multicolumn{1} {|c}{\textbf{Photometer}} & \multicolumn{1} {|c}{\textbf{Bandwidth (if relevant)}} & \multicolumn{1}{|c}{\textbf{FOV}} & \multicolumn{1}{|c|}{\textbf{Frequency}} & \multicolumn{1}{c|}{\textbf{Inclination}} \\

\endfirsthead

\multicolumn{1}{|c}{\textbf{Mission}} & \multicolumn{1} {|c}{\textbf{Photometer}} & \multicolumn{1} {|c}{\textbf{Bandwidth (if relevant)}} & \multicolumn{1}{|c}{\textbf{FOV}} & \multicolumn{1}{|c|}{\textbf{Frequency}} & \multicolumn{1}{c|}{\textbf{Inclination}} \\
\hline
\endhead

\hline

& SP1: 150~-~290~nm & - & 20 deg (H) $\times$ 5 deg (V) & 10~kHz & Limb \\ 
& SP2: 337~nm & 5.6~nm & 20 deg (H) $\times$ 5 deg (V) & 10~kHz & Limb \\ 
& SP3: 391~nm & 4.2~nm & 20 deg (H) $\times$ 5 deg (V) & 10~kHz & Limb \\ 
ISUAL & SP4: 608.9~-~753.4~nm & - & 20 deg (H) $\times$ 5 deg (V) & 10~kHz & Limb \\ 
& SP5: 777.4~nm & -  & 20 deg (H) $\times$ 5 deg (V) & 10~kHz & Limb \\ 
& SP6: 250~-~390~nm & - & 20 deg (H) $\times$ 5 deg (V) & 10~kHz & Limb \\ 
& AP1 (16 CH): 370~-~450~nm & - & 20 deg (H) $\times$ 5 deg (V) & 0.2, 2 or 20~kHz & Limb \\ 
& AP2 (16 CH): 530~-~650~nm & - & 20 deg (H) $\times$ 5 deg (V) & 0.2, 2 or 20~kHz & Limb \\ 
\hline
& PH1: 150~-~280~nm & - & 42.7$^{\circ}$ & 20~kHz & Nadir \\ 
& PH2: 332~-~342~nm & - & 42.7$^{\circ}$ & 20~kHz & Nadir \\ 
GLIMS & PH3: 755~-~766~nm & - & 42.7$^{\circ}$ & 20~kHz & Nadir \\ 
& PH4: 599~-~900~nm & - & 86.8$^{\circ}$ & 20~kHz & Nadir \\ 
& PH5: 310~-~321~nm & - & 42.7$^{\circ}$ & 20~kHz & Nadir \\ 
& PH6: 386~-~397~nm & - & 42.7$^{\circ}$ & 20~kHz & Nadir \\ 
\hline
& PH1: 145~-~230~nm & - & 61.4$^{\circ}$ & 100~kHz & Nadir \\ 
ASIM & PH2: 337~nm & 5~nm & 61.4$^{\circ}$ & 100~kHz & Nadir \\ 
& PH3: 777.4~nm & 5~nm & 61.4$^{\circ}$ & 100~kHz & Nadir \\ 
\hline
& PH1: 145~-~280~nm & - & 55$^{\circ}$ & 20~kHz & Nadir \\ 
TARANIS & PH2: 332 - 342 ~nm & - & 55$^{\circ}$ & 20~kHz & Nadir \\ 
& PH3: 757 - 765~nm & - & 55$^{\circ}$ & 20~kHz & Nadir \\ 
& PH4: 600~-~800~nm & - & 100$^{\circ}$ & 20~kHz & Nadir \\ 

\hline

\caption{Optical characteristic of the photometers onboard ISUAL \citep{Chern2003/JASTP}, GLIMS \citep{sato2015overview,Adachi2016/JASTP}, ASIM \citep{Neubert2006/ILWS} and TARANIS \citep{Blanc2007/AdSpR}. The observation mode (limb or nadir) of each photometer is indicated.} \label{tab:missions} \\

\end{longtable}
\normalsize

The photons emitted by TLEs can travel from the source to a nadir-pointing photometer without suffering an important atmospheric absorption. However, the signal of TLEs observed in the nadir can be contaminated by photons emitted by the parent-lightning discharge. An exception to this are the optical emissions in the LBH, as the photons emitted by lightning in these short wavelengths are totally absorbed by the atmosphere \citep{Mende2005/JGRA}. For this reason, the signals detected in the FPS(3,0), SPS(0,0) or FNS(0,0) cannot be analyzed following our inversion methods unless the parent lightning is out of the FOV.

In addition, recorded signals from TLEs taking place behind the limb would not be contaminated by the photons emitted by their lightning stroke. If the absorption of the atmosphere is used to correct the observed signals, it is then possible to apply our inversion methods to the analysis of behind-the-limb TLEs.

\subsubsection{Deduction of the reduced electric field in a halo reported by ISUAL}

\begin{figure}
\centering
\includegraphics[width=11cm]{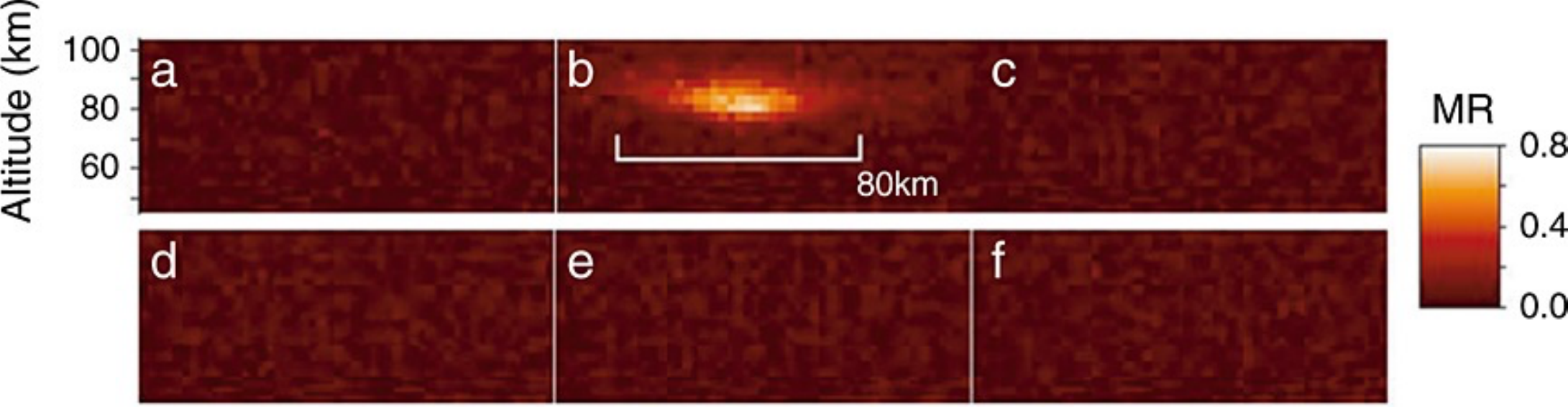}
\caption{Halo recorded by ISUAL at universal time 0626:38.806 on 31 July, 2006. The exposure time of each frame (a-f) is $\sim$30 ms. Image adapted from  \cite{Kuo2013/JGRA}.}
\label{fig:halokuo}
\end{figure}

\begin{figure}
\centering
\includegraphics[width=11cm]{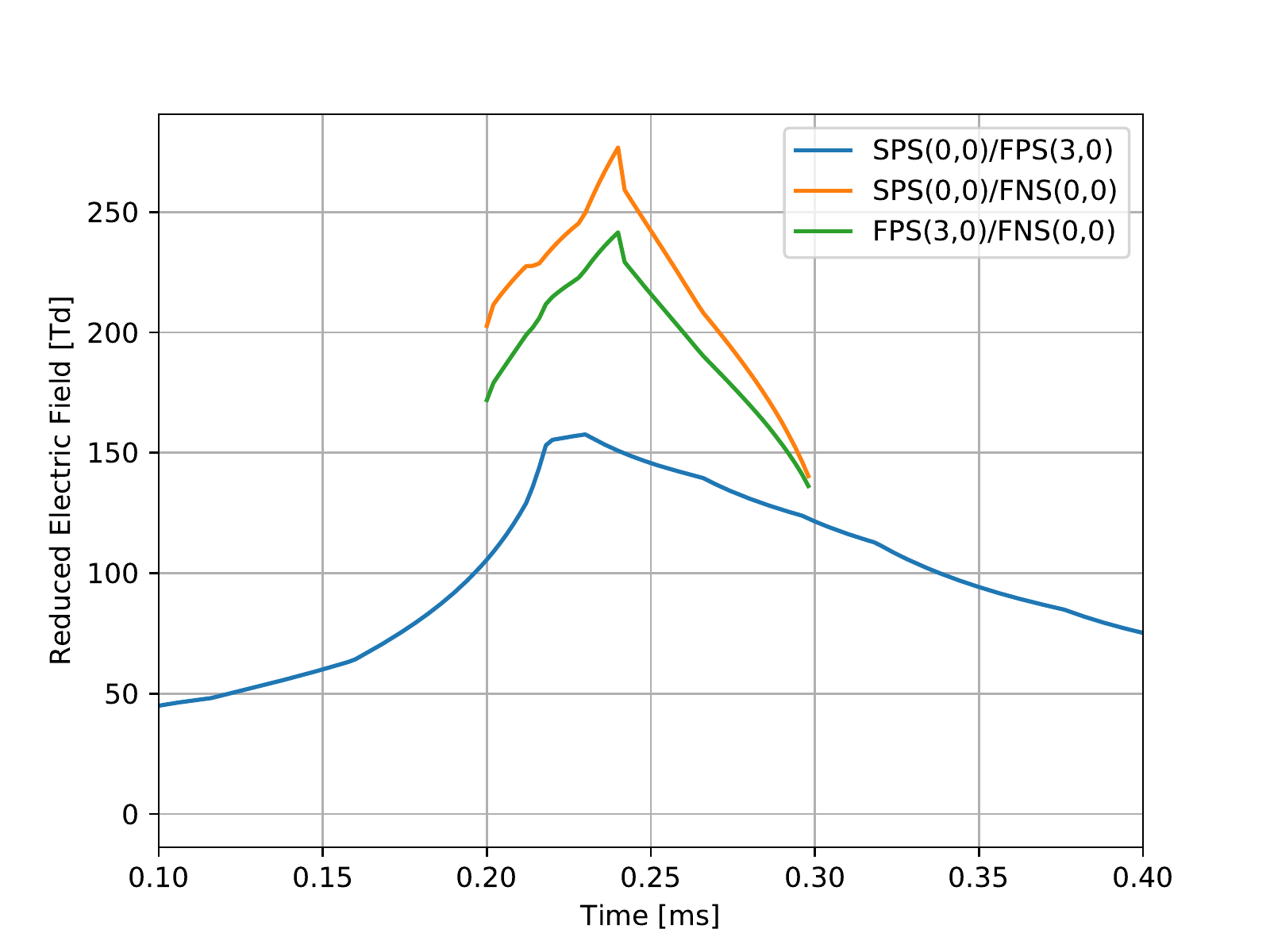}
\caption{Temporal evolution of the reduced electric field in the halo without no visible sprite reported by \cite{Kuo2013/JGRA} from the ratios of FPS(3,0) to SPS(0,0), SPS(0,0) to FNS(0,0) and FPS(3,0) to FNS(0,0). Optical emissions in the FNS(0,0) before 0.2~ms and after 0.3~ms cannot be used as a consequence of the signal to noise ratio. The value of the reduced electric field reported in \cite{Kuo2013/JGRA} ranged between 275~Td and 325~Td. We have used the VDF given by expression~(\ref{VDFC3P}).}
\label{fig:Ered}
\end{figure}

\cite{Kuo2013/JGRA} investigated a halo without a visible sprite reported by ISUAL on 31 July, 2006 (see figure~\ref{fig:halokuo}). The parent lightning was below the limb, at a distance of about 4100~km. Therefore, the photometric recordings were not contaminated by possible optical emissions from lightning.
\cite{Kuo2013/JGRA} estimated the maximum reduced electric field in this halo using the ratio of FNS(0,0) of N$_2^+$ to SPS(0,0) of N$_2$, obtaining a value ranging between 275~Td and 325~Td. In this section, we analyze the reported intensities of the ISUAL photometers SP2, SP3 and SP4 to estimate the reduced electric field using the methods developed in subsection~\ref{sec:opticalanalysis}. As proposed by \cite{Kuo2013/JGRA}, we calculate the emissions in SPS(0,0) of N$_2$, FNS(0,0) of N$_2^+$ and FPS(3,0) of N$_2$ from the SP2, SP3 and SP4 photometric recordings, respectively. We do that by considering the percentages of the respective band emissions that fit into each photometer, the atmospheric transmittances and the blueward shifts as proposed by \cite{Kuo2013/JGRA}.
We plot in figure~\ref{fig:Ered} the resulting reduced electric fields using the ratios of FPS(3,0) to SPS(0,0), SPS(0,0) to FNS(0,0) and FPS(3,0) to FNS(0,0). The value of the maximum reduced electric field calculated from the emission ratio of SPS(0,0) to FNS(0,0) agrees with the value obtained by \cite{Kuo2013/JGRA}. The reduced electric field calculated from the emission ratio of FPS(3,0) to FNS(0,0) is slightly below that value, while the field obtained from the ratio of FPS(3,0) to SPS(0,0) is about a factor of 2 below. The reason of the underestimation of the reduced electric field using the ratio of SPS(0,0) to FPS(3,0) is that this ratio does not depend significantly on the electric field when it is high (above 150~Td), as explained in subsection~\ref{sec:signalscomparison}. The reduced electric field obtained from the FNS(0,0) is only shown between 0.2~ms and 0.26~ms, because the signal in the FNS(0,0) is very noisy out of that range.

\subsubsection{Deduction of the source emissions of an elve reported by GLIMS}
\label{sourceGLIMS}

We apply the inversion method described in subsection~\ref{inversion} to the LBH signal of an elve reported by GLIMS at 16.28.04 (UT) on December 13, 2012. At the moment of the detection, the instrument was located at 422~km of altitude.

The optical camera Lightning and Sprite Imager (LSI) onboard GLIMS has a FOV of 28.3$^{\circ}$ and is equipped with two frequency filters, a broadband (768~nm~-~830~nm) filter (LSI-1) and a narrowband (760~nm~-~775~nm) filter (LSI-2).

Before applying the inversion method to the elve LBH signal, we have to deduce the horizontal separation between the photometers and the center of the elve ($l$) (see figure~\ref{fig:geometry}). For this purpose, we use the images taken by the LSI camera onboard GLIMS. Knowing the altitude (422~km) of GLIMS at the moment of the elve detection, assuming that the elve took place at an altitude of 88~km and knowing that the FOV of the camera is 28.3$^{\circ}$, the LSI camera can observe a square with a lateral dimension of 168~km in the plane of the elve (at 88~km of altitude). If we assume that the parent-lightning is located just below the center of the elve, we can use this information together with the number of pixels between the center of the camera FOV and the parent-lightning to calculate the horizontal separation between the elve center and the photometer. In this case, this separation is 55~km.

\begin{figure}
\centering
\includegraphics[width=13cm]{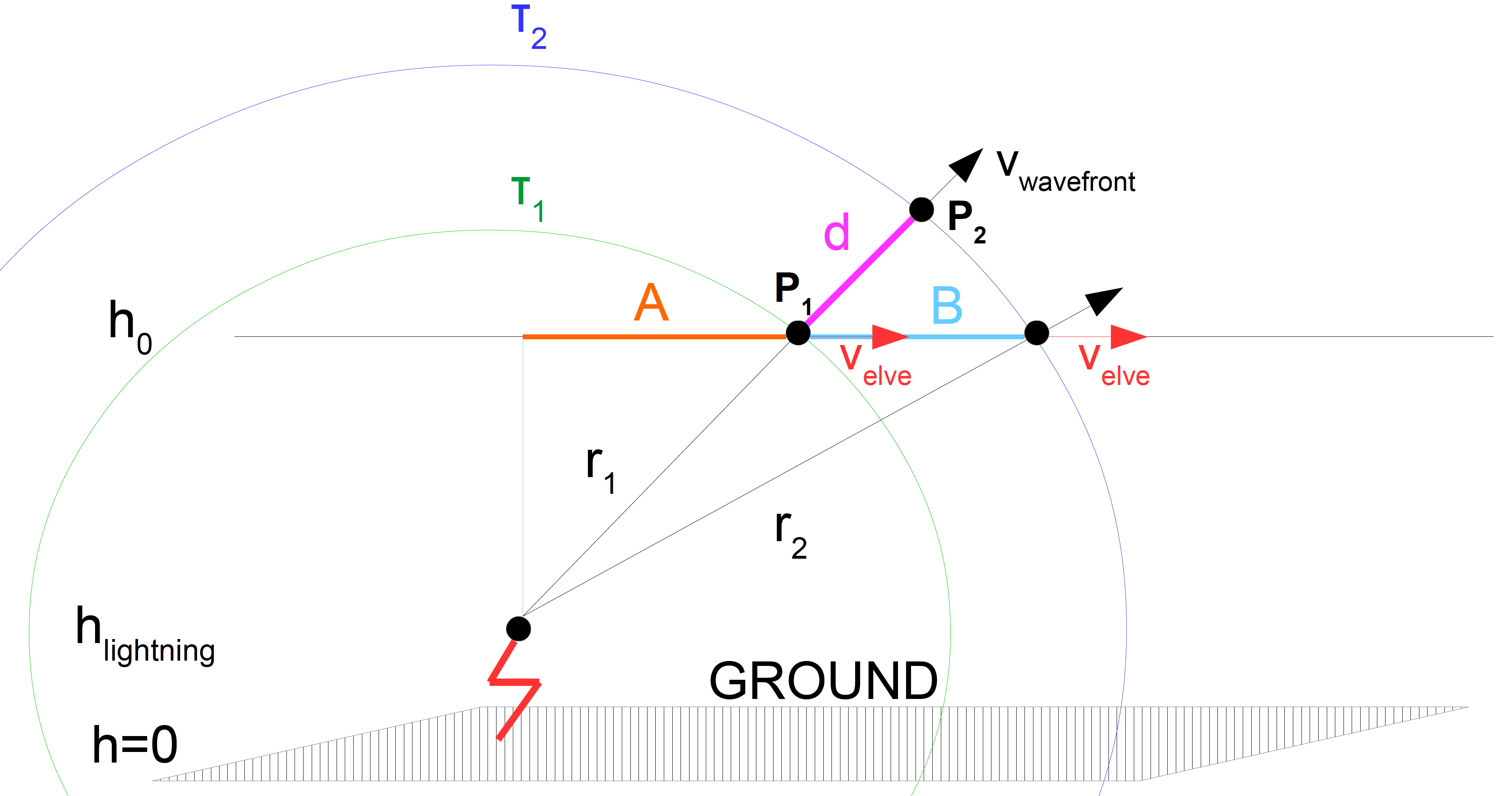}
\caption{Expansion of the elve wavefront and the elve radius during a time of $\tau$~=~$\tau_2$~-~$\tau_1$. Lightning and elve are located at altitudes of $h_{lightning}$ and $h_0$, respectively. The elve wavefront propagates at the velocity of light.}
\label{fig:elve_velocity}
\end{figure}

We also need to estimate the moment at which the source started its emissions in relation to the moment of detection of the first photon ($t_E$). That is, the difference between the time of the elve detection and the time of the elve onset. The deduction of this time is not direct, as the radius of the elve expands faster than the speed of light as can be demonstrated using the scheme plotted figure~\ref{fig:elve_velocity}.

We can calculate the expansion of the elve radius during a time~$\tau$~=~$\tau_2$~-~$\tau_1$ knowing that the pulse radiated by the lightning discharge travels at the speed of light and that the difference of altitudes between the elve and the lightning stroke is $h_0 - h_{lightning}$. At the moment $\tau_1$, the distance $r_1$ between stroke and the intersection of the wavefront with the ionosphere is given by

\begin{linenomath*}
\begin{equation}
r_1 =c \tau_1, \label{r1}
\end{equation}
\end{linenomath*}

then we can write the distance $A$, corresponding to the radius of the elve at a time of $\tau_1$, as

\begin{linenomath*}
\begin{equation}
A =\sqrt{c^{2}\tau_1^2 - (h_0 - h_{lightning})^2}. \label{A}
\end{equation}
\end{linenomath*}

If now we call $B$ the radius of the elve at a time $\tau_2 > \tau_1$, we can calculate the expansion of the elve radius during the time $\tau_2 - \tau_1$ as

\begin{linenomath*}
\begin{equation}
B - A =\sqrt{c^{2}\tau_1^2 - (h_0 - h_{lightning})^2} - \sqrt{c^{2}\tau_2^2 - (h_0 - h_{lightning})^2} > d, \label{B-A}
\end{equation}
\end{linenomath*}

while the elve wavefront would have advanced a distance of only $d = c(\tau_2 - \tau_1)$. 

Therefore, the first observed photon would be the one that travels the minimum path from the source to the photometer. We can calculate the time of flight of that photon by minimizing equation~(\ref{s}), obtaining the minimum optical path ($s_{min}$). The time at which $s(t) = s_{min}$ is the time after the elve onset at which the photon was emitted ($t_{min}$). The time of flight of that photon is then given by $\frac{s_{min}}{c}$. We can finally estimate the moment at which the source started its emissions in relation with the moment of detection of the first photon as the sum of the time of flight of the first detected photon and the time at which it was emitted as

\begin{linenomath*}
\begin{equation}
t_E =\frac{s_{min}}{c} + t_{min}. \label{te}
\end{equation}
\end{linenomath*}

\begin{figure}
\centering
\includegraphics[width=11cm]{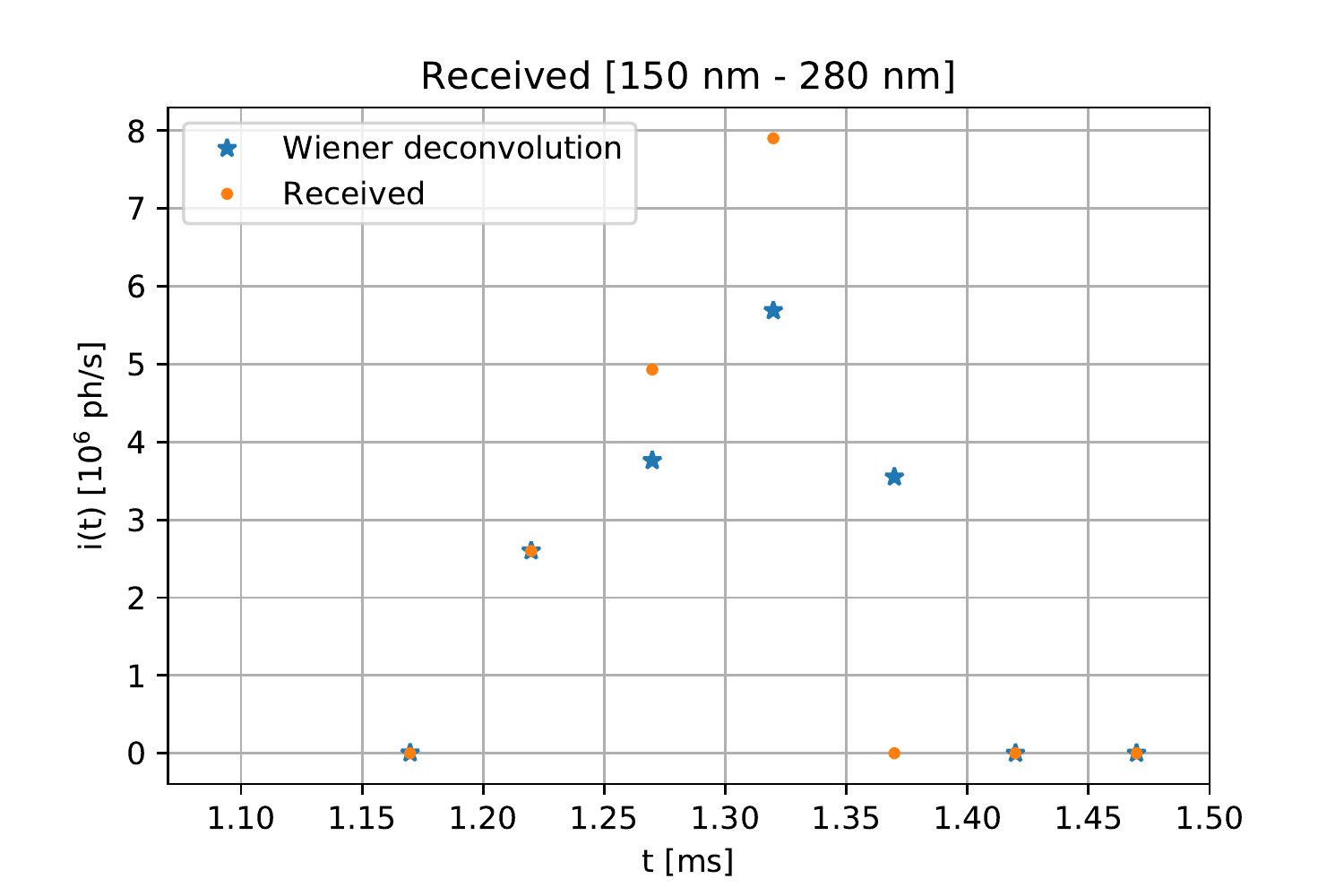}
\caption{Photons received by GLIMS in the 150~nm - 280~nm band from an elve (orange dots). Signal after the application of the Wiener deconvolution in order to approximate the elve as a thin ring (blue asterisk).}
\label{fig:I_decon_a2012-12-13_16280487397}
\end{figure}

The last step before applying the inversion method is to convert the detected signal from $W/m^2$ to photons/s. To do that, we multiply the signal by the area of the detector of radius 12~mm. We also divide by the energy of the received photons. The photons received by the photometer PH1 have wavelengths between 150~nm and 280~nm. As most of the photons emitted by the elve have wavelengths closer to 150~nm rather than to 280~nm \citep{Gordillo-Vazquez2010/JGRA, PerezInvernon2018/JGR}, we can calculate this energy as the corresponding energy of a photon with a wavelength of about 180~nm. The observed signal in photons/s is shown in figure~\ref{fig:I_decon_a2012-12-13_16280487397}. We also show in figure~\ref{fig:I_decon_a2012-12-13_16280487397} the signal obtained after applying the Wiener deconvolution.

\begin{figure}
\centering
\includegraphics[width=11cm]{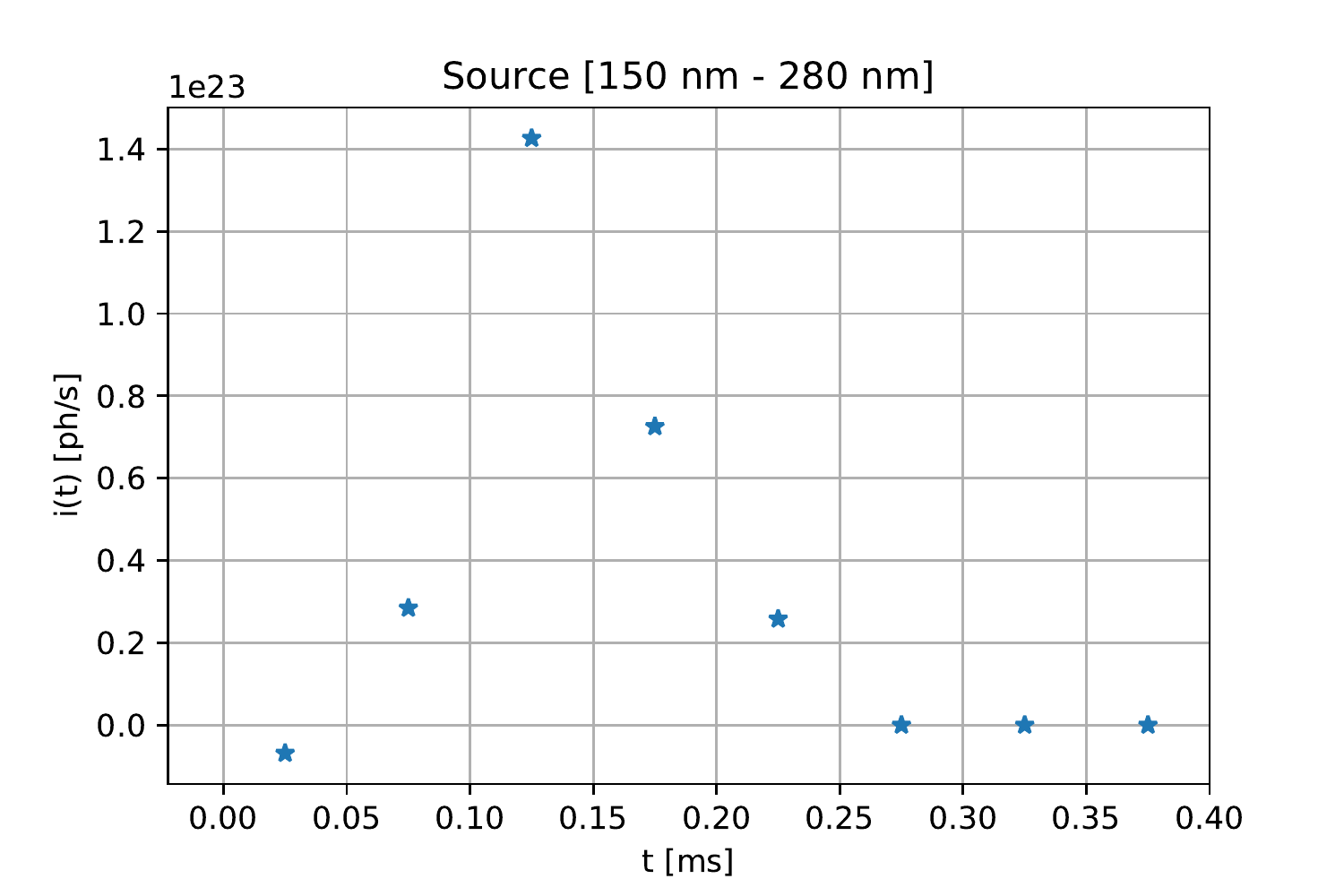}
\caption{Emitting source for an instantaneous thin ring.}
\label{fig:F_a2012-12-13_16280487397}
\end{figure}

Then, we can apply the Hanson method to the (blue) signal of figure~\ref{fig:I_decon_a2012-12-13_16280487397} in order to obtain the temporal evolution of the signal emitted by a thin ring. We plot the source optical emission in figure~\ref{fig:F_a2012-12-13_16280487397}.

\begin{figure}
\centering
\includegraphics[width=11cm]{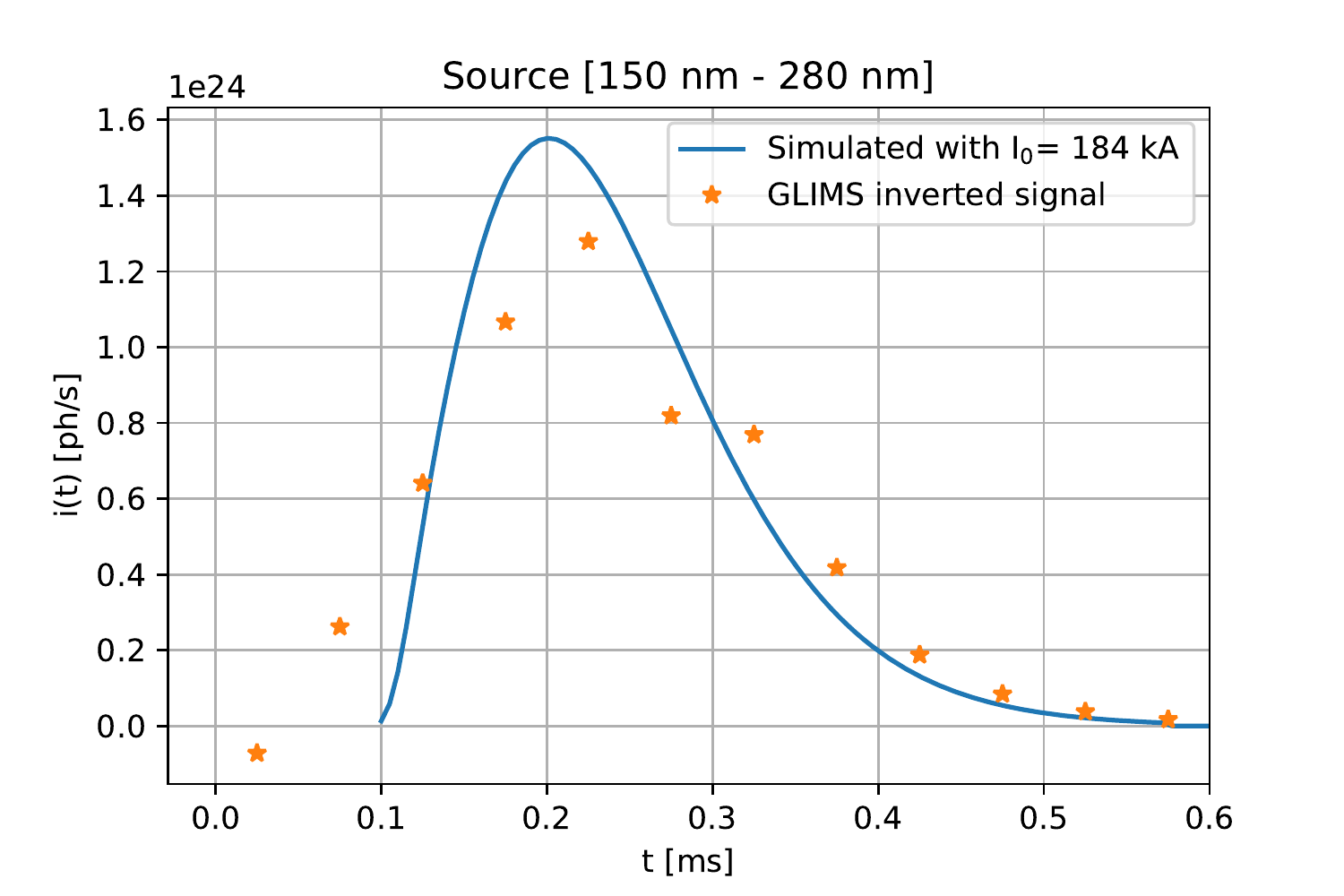}
\caption{Source plotted in figure~\ref{fig:F_a2012-12-13_16280487397} after the convolution with the corresponding decay function in order to obtain the emitting source (orange asterisk). This source emits photons with wavelengths between 150~nm and 280~nm. We also plot the simulated intensity emitted by an elve triggered by a CG lightning discharge with a current peak of 184~kA, a rise time of \SI{10}{\micro\second} and a total time of 0.5~ms. The simulated emitted intensity has been calculated using a FDTD elve model.}
\label{fig:comparison_GLIMS_FDTD}
\end{figure}

However, the obtained emitting source corresponds to the source of a thin ring. Therefore, we can obtain the real emitting source by convolving the obtained emitting source of an instantaneous thin ring with its corresponding decay function. The final emitting source is shown in figure~\ref{fig:comparison_GLIMS_FDTD}, together with the simulated intensity emitted by an elve triggered by a CG lightning discharge with current peak of 184~kA, a rise time of \SI{10}{\micro\second} and a total time of 0.5~ms.

\subsubsection{Remarks about the possibility of applying the inversion method to signals reported by other space missions}

In section~\ref{sourceGLIMS} we have applied the inversion method previously described in subsection~\ref{inversion} to an elve signal reported by GLIMS in a range of wavelengths between 150~nm and 280~nm. We mentioned that, in general, it is not possible to apply this method to the optical signals detected in other wavelengths, as the lightning flash would contaminate them. However, the detection of an elve whose parent lightning is out of the photometer FOV could be analyzed with our method providing that the position of the lightning stroke is known. A lightning detection network (local or global) could provide this information. In order to estimate the electric fields in halos, it would also be necessary to limit the analysis to halos whose parent lightning is outside the FOV.

In the case of ASIM and TARANIS, the FOV of the photometers and the optical cameras are the same. Therefore, if an elve is detected without its parent-lightning, the image of the elve taken by the optical cameras could be useful to estimate the position of the elve center and that of the lightning. In the case of ASIM, the photometers and the optical cameras are part of the Modular Multispectral Imaging Array (MMIA). The cameras record at 12~fps. However, the weak intensity of elves observed at the nadir makes difficult their detection by MMIA cameras. If the optical cameras of ASIM and TARANIS are not sensible enough to distinguish the shape of the elves, some detection network, as it was done in the case of GLIMS, must give the position of the lightning stroke.

The case of ISUAL is different, as it is capable of reporting elves taking place behind the limb without the contamination of the parent lightning. However, our inversion method is highly dependent of the elve position. \cite{Kuo2007/JGRA} developed a procedure to obtain the distance between ISUAL and the elve center by counting pixels in the image recorded by ISUAL optical cameras. However, we think that this procedure is not accurate enough to be used together with our inversion method. Therefore, a lightning detection network must probably give the position of the elve center, determined by the location of the parent lightning.

Finally, as the bandwidth of the ASIM and TARANIS photometers are quite similar to the bandwidth of the ISUAL photometers, we propose following the method of \cite{Kuo2013/JGRA} to calculate possible overlaps between different bands.

\section{Conclusions}
\label{sec:conclusions}

We have developed a general method to estimate the reduced electric field in upper atmospheric TLEs from space-based photometers. The first step of the proposed method uses of the observed emission ratio of two different spectral bands together with the continuity equation of the emitting species in order to estimate the rate of production by electron impact of the considered species. Then, the obtained ratio has been compared with the theoretical electric field dependent value calculated with BOLSIG+ \citep{Hagelaar2005/PSST}. Finally, we have calculated the reduced electric field that matches the observed emission ratio with the theoretical one.

The recorded optical signals of elves do not have the same temporal evolution as the emitting source. Therefore, we have developed an inversion procedure to calculate the emitting source of elves before applying the method to estimate the reduced electric field. However, this procedure is exclusively valid if the optical signal produced by the parent lightning does not contaminate the optical signal of the elve. We have successfully applied this inversion method to the optical signal emitted by an elve within the LBH band and recorded by GLIMS.

We have firstly applied this procedure to the predicted emissions of simulated halos and elves. In the case of a halo with a maximum electric field of 140~Td, we have estimated the reduced electric field using the ratio of FPS(3,0) to SPS(0,0), FPS(3,0) to FNS(0,0) and SPS(0,0) to FNS(0,0). The obtained reduced electric field overestimates the maximum field given by the model by less than 10\%. The LBH band of N$_2$ is not considered in the case of halos because this TLE descends through a region of the atmosphere where the quenching altitude of the vibrational levels of N$_2$(a$^1$ $\Pi _g$ , v = 0, ..., 12) by N$_2$ and O$_2$ changes rapidly.
In the case of a simulated elve with a maximum electric field of 210~Td, we have estimated the reduced electric field using the ratios of FPS(3,0) to SPS(0,0), FPS(0,0) to FNS(0,0), SPS(0,0) to FNS(0,0), FPS(3,0) to LBH, SPS(0,0) to LBH and FNS(0,0) to LBH. We have obtained a good agreement between the field given by the model and the estimated field using the observed ratios that include the FNS(0,0). The use of the ratio of FPS(3,0) to SPS(0,0) underestimates the value of the reduced electric field in about 15~\%. Finally, we have concluded that the ratios of FPS(3,0) to LBH and SPS(0,0) to LBH are not adequate to deduce the electric field in elves, as the results differ in more than 50~\% with respect to the expected quantity.

The ratio of SPS(0,0) to FNS(0,0) has been widely used to estimate reduced electric fields. We have found that the ratio of FPS(3,0) to FNS(0,0) leads to reduced electric field values that agree with those obtained by the SPS(0,0) to FNS(0,0) ratio. The application of our method to deduce the electric field of a halo reported by ISUAL \citep{Kuo2013/JGRA} has confirmed that the ratio of SPS(0,0) to FPS(3,0) emissions is not reliable if the electric field is below 150~Td or 200~Td.

\appendix

\section{Solution of the Fredholm integral equations of the first kind}
\label{ap:A}
This Appendix details the solution method of equation~(\ref{Itaus0}).
Following the notation used in \cite{Hanson1971/SIAM}, the kernel $k(s_i, t_j)$ of our integral equation is given by equation (\ref{kernel}), where the quantities $s_j$ and $t_j$ correspond to emissions and times of observation, respectively. In our case, the experimentally sampled function $g(s_i)$ is determined by the emitted photons per second from one ring, obtained after applying the Wiener deconvolution to the observed data $I(\tau)$ according to equation (\ref{Ihat}). 

Let us describe the method proposed by \cite{Hanson1971/SIAM} to solve a Fredholm integral equation of the first kind and its application to our particular case. We write the integral equation~(\ref{Itaus0}) as a linear system given by

\begin{linenomath*}
\begin{equation}
K F = G, \label{linear}
\end{equation}
\end{linenomath*}

where $G$ is a vector of size $m$ containing the observed signal, $F$ is an unknown vector of size $n$ corresponding to the emitted signal at the source, and $K$ is a $m \times n$ matrix representing the kernel $k(s_i, t_j)$. We assume that the measurements $g_i$ have some random error $\epsilon_i$ as

\begin{linenomath*}
\begin{equation}
g_i = \hat{g}_i + \epsilon_i, \label{giep}
\end{equation}
\end{linenomath*}

where $\mathrm{Var}(\epsilon_i) = \sigma_i^2$ and $\hat{g}_i$ represents the hypothetical measurements without error. In our case, we assume that the signal follows a Poisson distribution and set

\begin{linenomath*}
\begin{equation}
\sigma_i = \frac{\sqrt{g_i \Delta t}}{\Delta t},
\end{equation}
\end{linenomath*}

where $\Delta t$ is the integration time of the signal.

The Poisson distribution tends to a Gaussian distribution if the number of photons is large enough, as in this case. Then, assuming Gaussian noise we want to find the $F$ that maximizes the likelihood

\begin{linenomath*}
\begin{equation}
L = \alpha \prod_{i=1}^{m} \exp\left(-{\frac{\left(\left(K F\right)_i - G_i\right)^2}{\sigma_i^2}}\right), \label{L}
\end{equation}
\end{linenomath*}

where $\alpha$ is a normalization factor. This is the same as minimizing

\begin{linenomath*}
\begin{equation}
-\log\left({L}\right) = \sum\limits_{i=1}^m \frac{\left(\left(K F\right)_i - G_i\right)^2}{\sigma_i^2} = \left|\Sigma K F - \Sigma G \right| ^2, \label{minimizing}
\end{equation}
\end{linenomath*}

where 

\begin{linenomath*}
\begin{equation}
\Sigma = \left( \begin{matrix}
\sigma_1^{-1} & 0 & ... & 0 \\
0 & \sigma_2^{-1} & ... & 0 \\
0 & 0 & ... &  ... \\
... & ... & ... &  \sigma_m^{-1} \end{matrix} \right).
\end{equation}
\end{linenomath*}

In principle we can just minimize (\ref{minimizing}) using least squares but in some cases we would be over fitting the noise in $G$. We want an estimate of the ``extra" error that we allow in (\ref{minimizing}) to avoid strong oscillations in $F$. For this we use the Singular-Value Decomposition (SVD):

\begin{linenomath*}
\begin{equation}
\Sigma K = U \left[ \begin{matrix} 
S \\
0
\end{matrix}
\right] V^T,
\end{equation}
\end{linenomath*}

where $U$ and $V$ are orthogonal( $U^TU = \mathbb{I}$ and $V^TV = \mathbb{I}$) and

\begin{linenomath*}
\begin{equation}
\left[ \begin{matrix} 
S \\
0
\end{matrix} \right]
= \left[ \begin{matrix} 
S_1 & 0 & 0 & 0 & ... \\
0 & S_2 & 0 & 0 & ... \\
... & ... & ... & ... & ... \\
0 & 0 & 0 & ... & S_n \\
... & ... & ... & ... & ... \\
0 & 0 & 0 & ... & 0 \\
\end{matrix}
\right],
\end{equation}
\end{linenomath*}

results in a matrix $n\times m$.

To see how the SVD is useful we note that:

\begin{enumerate}
\item Since $U$ is orthogonal, multiplying by $U^T$ preserves the Euclidean norm, so $\left| \Sigma K F - \Sigma G \right| = \left| \left[ \begin{matrix} S \\ 0 \end{matrix} \right] V^T F - U^T \Sigma G \right|$
\item The column vectors of $U$ and $V$ form basis in $\mathbb{R}^m$ and $\mathbb{R}^n$, respectively. That is, calling these vectors $u_i$ and $v_i$ we have 
$u_i · u_j = \delta_{ij}$, $v_i · v_j = \delta_{ij}$.

so we can decompose
\begin{linenomath*}
\begin{equation}
F = x_1 v_1 + x_2 v_2 + ... \label{functionF}
\end{equation}
\begin{equation}
\Sigma G = e_1 u_1 + e_2 u_2 + ...
\end{equation}
\end{linenomath*}

this is,

\begin{linenomath*}
\begin{equation}
U^T \Sigma G = \left[ \begin{matrix} e_1 \\ e_2 \\ ... \\ e_m \end{matrix} \right], \label{UV1}
\end{equation}
\begin{equation}
V^T F = \left[ \begin{matrix} x_1 \\ x_2 \\ ... \\ x_n \end{matrix} \right]. \label{UV2}
\end{equation}
\end{linenomath*}

But then we have the following vector with $m$ components

\begin{linenomath*}
\begin{equation}
\left[ \begin{matrix} S \\ 0 \end{matrix} \right] V^T F = \left[ \begin{matrix} s_1 x_1 \\ s_2 x_2 \\ ... \\ s_n x_n \\ 0 \\ ... \end{matrix} \right] . \label{S0}
\end{equation}
\end{linenomath*}

Combining equations~(\ref{UV1}), (\ref{UV2}) and~(\ref{S0}), we have that 

\begin{linenomath*}
\begin{equation}
\left[ \begin{matrix} S \\ 0 \end{matrix} \right] V^T F - U^T \Sigma G = \left[ \begin{matrix} (s_1 x_1 - e_1) \\ (s_2 x_2 - e_2) \\ ... \\ (s_n x_n - e_n) \\ e_{n+1} \\ ... \\ e_m \end{matrix} \right],
\end{equation}
\end{linenomath*}

and the norm that we want to ``minimize" is

\begin{linenomath*}
\begin{equation}
\left| \Sigma K F - \Sigma G \right|^2 = \sum\limits_{i=1}^m\left( s_i x_i - e_i \right)^2 + \sum\limits_{i=n+1}^m e_i^2. \label{varsigma}
\end{equation}
\end{linenomath*}

We minimize equation~(\ref{varsigma}) by setting $x_i = e_i / s_i $. But, as said above, we do not want to minimize this ``too much", we accept an error

\begin{linenomath*}
\begin{equation}
\left| \Sigma K F - \Sigma G \right|^2 \simeq \mathrm{Var}(|\Sigma G|) = \frac{\mathrm{Var}(g_1)}{\sigma_1^2} + \frac{\mathrm{Var}(g_2)}{\sigma_2^2} + ... = m.
\end{equation}
\end{linenomath*}

So we set as many $x_i = 0$ as we can by dropping first those with the smallest $s_i$ because they are responsible of the largest oscillations in $F$. That is, we find the smallest $q$ that satisfies

\begin{linenomath*}
\begin{equation}
\sum\limits_{i=q+1}^m e_i^2 < m, \label{eq}
\end{equation}
\end{linenomath*}

and calculate all the $x_i = \frac{e_i}{s_i}$ with $i$ ranging between 1 and $q$.  Once we have all the $x_i$ we use equation~(\ref{functionF}) to get $F$.

\end{enumerate}

\section*{Acknowledgement}

This work was supported by the Spanish Ministry of Science and Innovation, MINECO under projects ESP2015-69909-C5-2-R and ESP2017-86263-C4-4-R and by the EU through the H2020 Science and Innovation with Thunderstorms (SAINT) project (Ref.~722337) and the FEDER program. FJPI acknowledges a PhD research contract, code BES-2014-069567. AL was supported by the European Research Council (ERC) under the European Union H2020 programme/ERC grant agreement 681257. TA was supported by JSPS KAKENHI Grant Number 24840040. The simulation data presented here are available from figshare repository at https://figshare.com/s/f1c9f6c7bc728d6669dd. JEM-GLIMS and ISUAL data presented in this study are available in the webpages http://www.ep.sci.hokudai.ac.jp/\textasciitilde{}jemglims/index-e.html and http://formosat.tw/nspo\_esok/, respectively. Alternatively, requests for data and codes used to generate or displayed in figures, graphs, plots, or tables may be made to the authors F.J.P.I (fjpi@iaa.es), A.L (aluque@iaa.es), or F.J.G.V (vazquez@iaa.es). 

\newcommand{\pra}{Phys. Rev. A} 
\newcommand{\jgr}{J. Geoph. Res. } 
\newcommand{\jcp}{J. Chem. Phys. } 
\newcommand{\ssr}{Space Sci. Rev.} 
\newcommand{\planss}{Plan. Spac. Sci.} 
\newcommand{\pre}{Phys. Rev. E} 
\newcommand{\nat}{Nature} 
\newcommand{\icarus}{Icarus} 
\newcommand{\ndash}{-}

\end{document}